%% file: Global_local_asynchrone_HAL.tex
\documentclass[11pt]{article}
\usepackage{amssymb}
\usepackage[utf8]{inputenc}
\usepackage{graphicx,color}
\usepackage[export]{adjustbox}
\usepackage[width=16cm,height=25cm]{geometry}
\usepackage{subcaption}
\usepackage{pgfplots}\pgfplotsset{compat=newest}
\usepackage{amsmath,amsfonts,amssymb,amsthm,mathabx}
\usepackage[mathscr]{euscript}  %Font mathscr
\usepackage{float}
\usepackage{array,multirow}
\usepackage{url}
\usepackage[numbers]{natbib}
\usepackage[inkscapearea=crop]{svg}
\makeatletter
\newif\if@restonecol
\makeatother

\usepackage[algo2e,ruled,lined,titlenumbered,commentsnumbered]{algorithm2e}
\usepackage{ulem}
\theoremstyle{remark}
\newtheorem{remark}{Remark}

\newcolumntype{L}[1]{>{\raggedright\let\newline\\\arraybackslash\hspace{0pt}}m{#1}}
\newcolumntype{C}[1]{>{\centering\let\newline\\\arraybackslash\hspace{0pt}}m{#1}}
\newcolumntype{R}[1]{>{\raggedleft\let\newline\\\arraybackslash\hspace{0pt}}m{#1}}
\newcommand{\address}[1]{\def\@address{#1}}
\title{Asynchronous global-local non-invasive coupling \\for linear elliptic problems}
\author{Ahmed El Kerim$^{1,3}$, Pierre Gosselet$^2$, Frédéric Magoulès$^{3,4}$, \\
\small $^1$ Universit\'e Paris-Saclay, ENS Paris-Saclay, CNRS, LMPS, \\\small ahmed.elkerim@ens-paris-saclay.fr\\
\small	$^2$ Université de Lille, CNRS, Centrale Lille / LaMcube, \\\small pierre.gosselet@univ-lille.fr\\
\small	$^3$ Universit\'e Paris-Saclay, CentraleSup\'elec / MICS, \\\small frederic.magoules@hotmail.com\\
\small	$^4$ Faculty of Engineering and Information Technology, University of Pécs\\
}
\begin{document}
	\maketitle

\newcommand{\dom}{\ensuremath{\Omega}}
\newcommand{\inter}{\ensuremath{\Gamma}}

\newcommand{\F}{\ensuremath{^F}}
\newcommand{\G}{\ensuremath{^G}}
\newcommand{\GT}{\ensuremath{^{G^T}}}
\newcommand{\Fs}{\ensuremath{^{s,F}}}
\newcommand{\Gs}{\ensuremath{^{s,G}}}
\newcommand{\Gsm}{\ensuremath{^{s,G^{-1}}}}
\newcommand{\Gz}{\ensuremath{^{0,G}}}
\newcommand{\Fz}{\ensuremath{^{0,F}}}
\newcommand{\Fsm}{\ensuremath{^{s,F^{-1}}}}

\newcommand{\Fss}{\ensuremath{^{s,F^*}}}
\newcommand{\FsT}{\ensuremath{^{s,F^T}}}
\newcommand{\Gss}{\ensuremath{^{s,G^*}}}
\newcommand{\GsT}{\ensuremath{^{s,G^T}}}
\newcommand{\GG}{\ensuremath{^{\Gamma,G}}}

\newcommand{\s}{\ensuremath{^{s}}}
\newcommand{\sstar}{\ensuremath{^{s^*}}}
\newcommand{\z}{\ensuremath{^{0}}}
\newcommand{\sT}{\ensuremath{^{s^T}}}
\newcommand{\zT}{\ensuremath{^{0^T}}}

\newcommand{\R}{\ensuremath{^R}}
\newcommand{\Rs}{\ensuremath{^{R^*}}}
\newcommand{\Gm}{\ensuremath{^{G^{-1}}}}

\newcommand{\bK}{\ensuremath{\mathbf{K}}}
\newcommand{\bA}{\ensuremath{\mathbf{A}}}
\newcommand{\bB}{\ensuremath{\mathbf{B}}}
\newcommand{\bI}{\ensuremath{\mathbf{I}}}
\newcommand{\bJ}{\ensuremath{\mathbf{J}}}
\newcommand{\bS}{\ensuremath{\mathbf{S}}}
\newcommand{\bT}{\ensuremath{\mathbf{T}}}
\newcommand{\bP}{\ensuremath{\mathbf{P}}}
\newcommand{\bM}{\ensuremath{\mathbf{M}}}
\newcommand{\bX}{\ensuremath{\mathbf{X}}}
\newcommand{\bY}{\ensuremath{\mathbf{Y}}}
\newcommand{\bL}{\ensuremath{\mathbf{L}}}

\newcommand{\bSnl}{\ensuremath{\mathcal{S}}}

\newcommand{\bu}{\ensuremath{\mathbf{u}}}
\newcommand{\bw}{\ensuremath{\mathbf{w}}}
\newcommand{\bp}{\ensuremath{\mathbf{p}_\Gamma}}
\newcommand{\bq}{\ensuremath{\mathbf{q}}}
\newcommand{\bv}{\ensuremath{\mathbf{v}}}
\newcommand{\br}{\ensuremath{\mathbf{r}}}
\newcommand{\bb}{\ensuremath{\mathbf{b}}}
\newcommand{\bx}{\ensuremath{\mathbf{x}}}
\newcommand{\by}{\ensuremath{\mathbf{y}}}
\newcommand{\bs}{\ensuremath{\mathbf{s}}}
\newcommand{\be}{\ensuremath{\mathbf{e}}}

\newcommand{\f}{\ensuremath{\mathbf{f}}}
\newcommand{\foint}{\ensuremath{\mathbf{f}_{int}}}
\newcommand{\foext}{\ensuremath{\mathbf{f}_{ext}}}
\newcommand{\foexti}{\ensuremath{\mathbf{f}_{ext,i}}}
\newcommand{\foextb}{\ensuremath{\mathbf{f}_{ext,b}}}

\newcommand{\lam}{\ensuremath{\boldsymbol{\lambda}}}

\newcommand{\actifs}{\underline{\mathcal{A}}}
\newcommand{\inactifs}{\mathcal{I}}
\newcommand{\bpu}{\mathbf{p}_{\Gamma,1}}
\newcommand{\bpd}{\mathbf{p}_{\Gamma,2}}

\newcommand{\hbp}{\ensuremath{\mathbf{\hat{p}}_\Gamma}}

\newcommand{\hbq}{\ensuremath{\mathbf{\hat{q}}}}
\newcommand{\hbb}{\ensuremath{\mathbf{\hat{b}}}}
\newcommand{\hbS}{\ensuremath{\mathbf{\hat{S}}}}
\newcommand{\hbY}{\ensuremath{\mathbf{\hat{Y}}}}
\newcommand{\tbv}{\ensuremath{\mathbf{\tilde{v}}}}

		\begin{abstract}
This paper presents the first asynchronous version of the Global/Local non-invasive coupling, capable of dealing efficiently with multiple, possibly adjacent, patches. We give a new interpretation of the coupling in terms of primal domain decomposition method, and we prove the convergence of the relaxed asynchronous iteration. The asynchronous paradigm lifts many bottlenecks of the Global/Local coupling performance. We illustrate the method on several linear elliptic problems as encountered in thermal and elasticity studies.
		\end{abstract}
	\section{Introduction}
	Engineering problems are often defined on very different scales, ranging from a coarse scale to model the whole structure to very fine scales that allow for the local details to be resolved. A method frequently used in the industry to link the scales is the submodeling \cite{kelley1982, ransom1992computational, cormier_Aggressive_1999}. This non-intrusive method is simple to implement but has shown limits regarding the accuracy of the results. 
	
	The non-invasive Global-Local coupling technique was first proposed and implemented in \cite{gendre2009non}. It aims at making submodeling accurate by means of iterations. It extends some previous reanalysis techniques \cite{jara-almonte_specified_1988, whitcomb_iterative_1991, whitcomb_application_1993}, and it has strong connections with Schwarz domain decomposition methods \cite{HECHT:2009:NZS, blanchard.2018.1} and multiscale methods \cite{Ladeveze1} while preserving the non-intrusive character of submodeling. Thus, it was implemented to couple research codes and legacy commercial software like Abaqus \cite{blanchard2019}, Code\_Aster \cite{duval2014non}, or Z-set \cite{Wangermez}.
	%The non-invasive local-global coupling technique proposed by Allix and first implemented in \cite{gendre2009non} is an iterative technique which aims at making accurate the well known submodeling method frequently used by industrialists \cite{kelley1982, ransom1992computational, cormier_Aggressive_1999}. It is strongly related to many reanalysis techniques \cite{jara-almonte_specified_1988, whitcomb_iterative_1991, whitcomb_application_1993} and domain decomposition methods \cite{HECHT:2009:NZS}. 
	
	The philosophy is to start from a simplified global model and then allow local alterations (geometry, material, load, and mesh) to be inserted and their effect to be evaluated without heavy intervention on the initial model (see~\cite{AllixGosselet} for a pedagogic presentation). It was successfully applied in many contexts like the introduction of local plasticity and geometrical refinements \cite{gendre2009non}, the computation of the propagation of cracks in a sound model \cite{duval2014non}, the evaluation of stochastic effects with deterministic computations \cite{safatly2012methode,nouy2017}, the taking into account of the exact geometry of connectors in an assembly of plates \cite{GUGUIN.2016.1}. In \cite{duval2014non} the method was used in order to implement a nonlinear domain decomposition method \cite{keyes1995aerodynamic,cresta2007nonlinear,hinojosa2014,NEGRELLO.2016.1} in a non-invasive manner in Code\_Aster. Extension of the approach to explicit dynamics was proposed in \cite{bettinotti2013coupling}, improved in \cite{bettinotti2014}  and applied to the prediction of delamination under impact loading in \cite{bettinotti2017}.
	
	All the above applications were developed in a synchronous framework that has been taken advantage of by accelerators (Aitken, quasi-Newton, Krylov), see \cite{blanchard.2018.1} where the method is proved to be an implementation of an alternating Dirichlet-Robin approach where the Robin parameter corresponds to the condensation of the coarse domain covered by the patch. However, due to the alternating nature of the method, its computational performance is inherently limited, with some processors idling while others are computing. This paper aims at deriving an asynchronous version of the global-local coupling, which enables us to get rid of most waiting periods.
	
	Asynchronous iteration was introduced in \cite{Chazan}, under the name of chaotic relaxation, to solve large linear systems. It has subsequently been the subject of several studies, \cite{Miellou} generalized the method to nonlinear problems, the work in \cite{Baudet} allowed the first implementation of asynchronous methods on multiprocessor architectures, in \cite{Tarazi, Bertsekas1983} convergence results for the asynchronous iterations based on the notion of classical contraction was presented, recent work in \cite{Asynch_Richardson} show interesting theoretical and practical results for the Richardson iterations from the asynchronous point of view.
	
	Several works have shown that domain decomposition methods are well suited for asynchronous parallel computation, such as alternating Schwarz \cite{Altern_async}, optimized Schwarz \cite{opti_shw_async, Yamazaki.19, Garay}, sub-structuring methods \cite{Magoules1, Gbikpi-Benissan2}, primal Schur domain decomposition method \cite{Gbikpi-Benissan2022} and also multigrid methods \cite{mutligrid_async}. In \cite{Survey_async, async_iters}, one can find a global review of asynchronous iterations from both theoretical and implementation points of view.\medskip
	
	Our study is conducted on linear elliptic problems discretized by the finite element approach. We prove the convergence of relaxed iterations using the theory of paracontractions \cite{paracontractions92}, and illustrate it on several examples of thermal and elasticity problems.
	
	The paper is organized as follows: in Section~\ref{sec:derivation} a new derivation of the method is proposed, in Section~\ref{sec:async} the asynchronous framework is exposed and studied, illustrations are given in Section~\ref{sec:applications}.
	
	%%%%%%%%%%%%%%%%%%%%%%%%%%%%%%%%%%%%%%%%%%%%%%%%%%%%%
	\section{The non-invasive global/local coupling}\label{sec:derivation}

\input{GLprincipLin}

	%%%%%%%%%%%%%%%%%%%%%%%%%%%%%%%%%%%%%%%%%%%%%%%%%%%%%

	\section{Asynchronous version}\label{sec:async} 
	\subsection{Introduction}
	In previous section, the Global/Local coupling has been presented as a robust and non-invasive method. However, from a performance point of view, it remains limited and less adapted to high performance computing, due to its alternating nature, see~\cite{blanchard.2018.1}. As an illustration, we consider the case of two zones of interest and a global problem as presented in Figure~\ref{fig:scenario}. 
	
\begin{figure}[H]\centering
		\begin{subfigure}{.45\textwidth}\centering
			\includegraphics[width=0.8\textwidth]{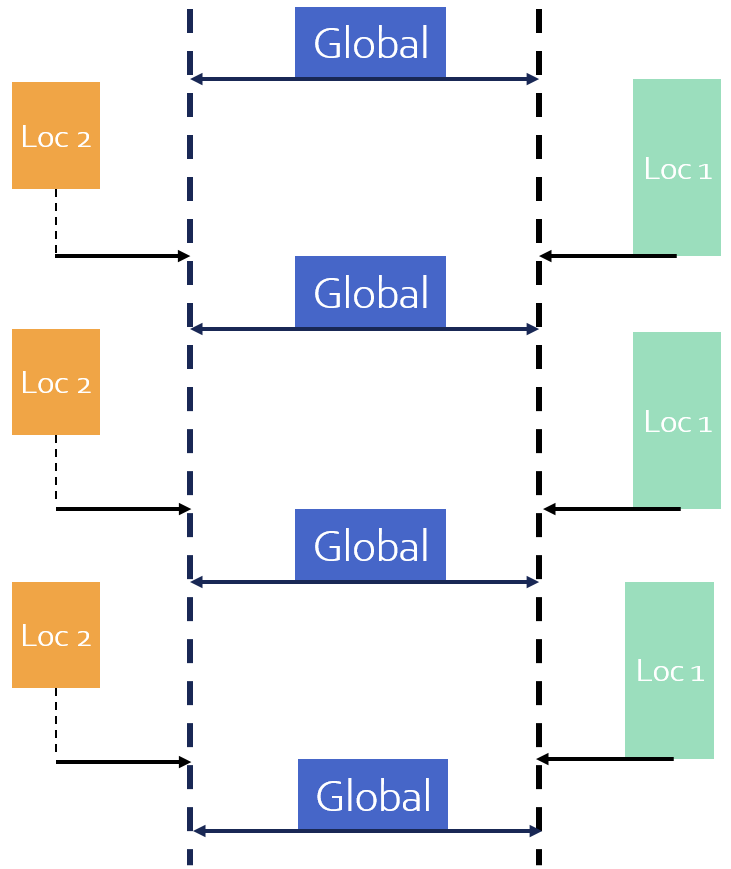}
			\caption{Synchronous iteration}
			\label{Synchronous model}
		\end{subfigure}
		%	\begin{minipage}{.45\textwidth}\centering
			%		\includegraphics[width=0.8\textwidth]{./GL_synchrone.png}
			%		\caption{Time course of the synchronous iteration}
			%		\label{Synchronous model}
			%	\end{minipage}
		%	\begin{minipage}{.08\textwidth}\centering
			\includegraphics[width=0.07\textwidth]{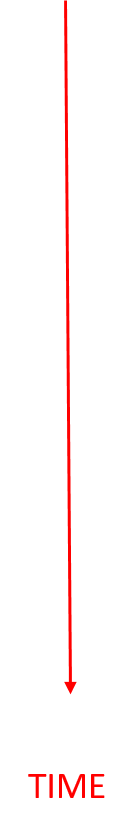}
			%	\end{minipage}
		\begin{subfigure}{.45\textwidth}\centering
			\includegraphics[width=0.8\textwidth]{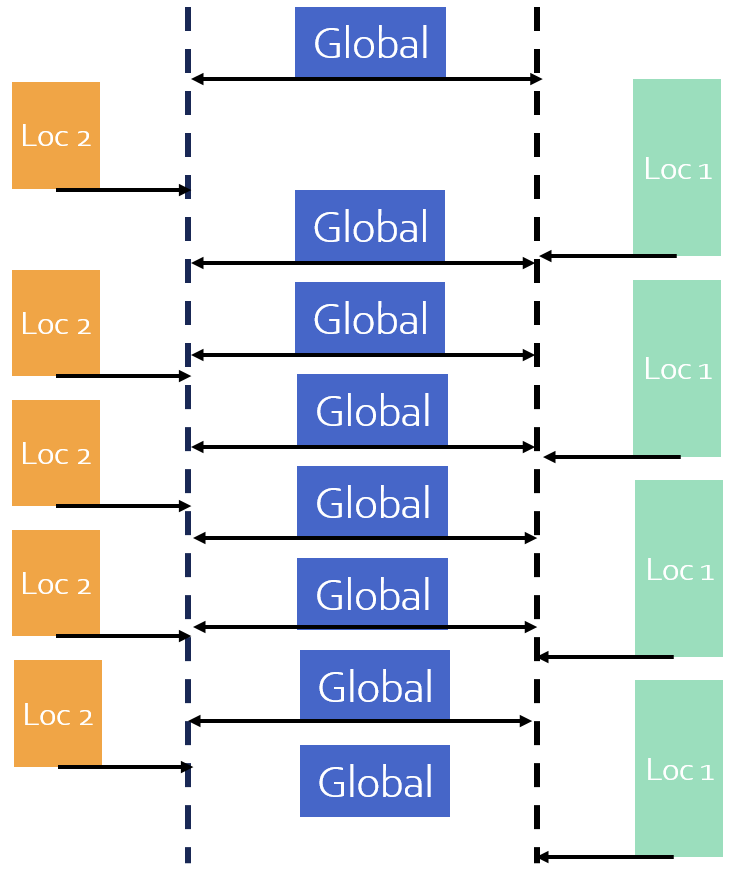}
			\caption{Asynchronous iteration}
			\label{Asynchronous with wait}
		\end{subfigure}
		\caption{Time course of the Global/Local coupling in the case of two patches}\label{fig:timecourse}
	\end{figure}

	Figure~\ref{Synchronous model} presents the time sequence of the classical synchronous approach, which alternates between global and parallel Fine calculations. Such an organization generates waiting and inactivity times on both sides, which seriously affects the performance. This phenomenon would be even amplified by bad load balancing, communication delays, or machine failures. 
	
	We establish an asynchronous parallel version of the Global/Local coupling to address these problems. 
	The idea is to allow each processor to work at its own pace without waiting for the other processors, considering only the latest version of the available data. This technique leads to the time sequence of Figure~\ref{Asynchronous with wait} where processors only wait when they have no new data to process
	
	Based on the figure\ref{Asynchronous with wait}, the algorithm \ref{alg:staAit}, presents an asynchronous version of the algorithm \ref{alg:staSit}.
	
	\begin{algorithm2e}[H]\caption{Asynchronous iterations}\label{alg:staAit}\DontPrintSemicolon
		Initialization $\bp=0$, $\omega$ sufficiently small\;
		\While{$\|\br\|$ is too large}{
			\If{Rank 0 is available and detects at least one new $\bq\s$} {
				Resolution of the Global system~\eqref{eq:global} or~\eqref{eq:discreteglo}, $\bu\G_\Gamma = \bS\Gm (\bp + \bb\G)$\;
				\If{$\Omega\z$ exists }{Post-processing~\eqref{eq:sd0}, $\bq\z:=\lam\z = \bS\z \bu\Gz_\Gamma - \bb\Gz $}
				Global \textit{scatters} $\bA\sT\bu\G_\Gamma$ to subdmains $s>0$\;
			}
			\For{$s>0$}{
				\If{Subdomain $s>0$ is available and detects new $(\bA\sT\bu\G)$}{
					Patch \textit{receives} $\bA\sT\bu\G_\Gamma$\;
					Fine solution~\eqref{eq:sdF}, $\lam\Fs = \bS\Fs\bJ\s\bA\sT\bu\G_\Gamma-\bb\Fs$\;
					Patch \textit{sends} of $\bq\s:=\bJ\sT\lam\Fs$ to the Global
				}
			}
			Global \textit{gathers} all $\bq\s$\;
			Global computes residual $\br=-\sum_s \bA\s\bq\s$\;
			Global updates $\bp = \bp + \omega\br$}
	\end{algorithm2e}
	%%One of the problems faced in the asynchronous switch is the detection of convergence of the method. In the literature, we find several approaches we mention some of them. [Savari et Besrstaks 1996] the idea is to make an exact evaluation of the residue from a snapshot of the general state of the solution. Another idea is to use the principle of the elected node like in a tree. It starts by checking a local convergence on each processor and then goes to the step of the global convergence verification. With the asynchronous version of the coupling, the residual keeps the mechanical meaning of the lack of balance between subdomains in the \textit{reference} problem, and anyhow it is updated as soon as one computation is complete. And consequently, in our situation, we don't need to implement a specific convergence detection protocol for the asynchronous version.

	%One crucial point compared to the synchronous version is that the reaction of the complement subdomain $\lam\Gz$ must be kept synchronized with the \textit{global} iteration.
	
	Note that the detection of the convergence of asynchronous iteration may require a specific, sometime complex, protocol~\cite{Stopping1, Stopping2}. Since the Global/Local coupling always assembles the residual on the Global model, our stopping criterion can be the same as in the synchronous case, simply based on the norm of the residual.
	
	% to do so, the simplest way is to implement formula~\eqref{eq:lambdaint}.

	% However, unlike the synchronous model, the asynchronous model is different from the sequential one, whose convergence study has been the subject of several works. As in Chazan's paper for linear problems based on algebraic properties of matrices, other works like those of Miellou and Scyeld have generalized the study for non-linear problems based on contraction operators.\\
	%In the following sub-section we detail our convergence proof by specifying the main reference of our wo

	\subsection{Convergence proof of the asynchronous iteration}
	Proving the convergence of asynchronous iteration can be tedious. In our case, we have the advantage of the Global domain playing a special role such that it can be used to cadence the solver. 
	Referring to Algorithm~\ref{alg:staAit}, we can consider that during the step from iteration $j$ to $j+1$,  some patches provide new pieces of information in order to evaluate the residual, anyhow these pieces of information may be related to old configurations \textbf{$p_{j-\sigma(s,j)}$} where \textbf{$\sigma(s,j)\geqslant0$} is a delay function. So that we can model the asynchronous iteration as: 
	\begin{equation}
		\begin{cases}
			\begin{aligned}
				&\bu\G_{\Gamma,j}= \bS\Gm(\bb\G+{\bp}_{j}) \\
				&\text{ If } s = 0: \bq\z_{j} = \bS\z(\bA\zT\bu\G_{\Gamma,j}-\bb\Gz)\\
				&\text{ If } s > 0: \bq\s_{j} =
				\begin{cases}
					\bJ\sT(\bS\Fs\bJ\s\bA\sT\bu\G_{\Gamma,j-\sigma(s,j)}-\bb\Fs) \text{ if updated }\\
					\qquad \bq\s_{j-1}, \text{ if not updated } \\
				\end{cases} \\
				&\br_{j}=-\left( \bA\z \bq\z_{j} + \sum_{s>0}\bA\s\bq\s_{j}\right) \\
				&{\bp}_{j+1} = {\bp}_{j}+ \omega\br_{j} 
			\end{aligned}
		\end{cases}
	\end{equation}
	For subdomains not updated, we set: $\sigma(s,j)=\sigma(s,j-1)+1$.
	
	It is crucial to note that if it exists, subdomain $0$ always contributes to the evaluation of the residual because computing $\bq\z_{j+1}$ is only a cheap postprocessing of the Global solution. In order to unify notations, we introduce $\sigma(0,j)=0$, $\forall j$, and then:
	\begin{equation}\label{eq:pwithsigma}
		\begin{aligned}
			{\bp}_{j+1}& = {\bp}_j -\omega\sum_{s=0}^N \bA\s \bJ\sT(\bS\Fs\bJ\s \bA\sT \bS\Gm ({\bp}_{j-\sigma(s,j)} + \bb\G)-\bb\Fs) \\
			&= {\bp}_j - \omega  \left(\sum_{s=0}^N \hbS\Fs \bS\Gm {\bp}_{j-\sigma(s,j)} + \hbb\right)
		\end{aligned}
	\end{equation}
	Note that this expression is valid only after all local patches have at least contributed once to the estimation of the residual. 
	
	In order to ensure that at some point all patches provide new information, we assume that:
	\begin{equation}
		\begin{aligned}
			&\exists D\geqslant 0 \text{ such that }\forall (s,j), \ \sigma(s,j)\leqslant D
		\end{aligned}
	\end{equation}
	For a given delay $0\leqslant k\leqslant D$, we write $\varpi(k,j)$ the set of subdomains $(s)$ such that $\sigma(s,j)=k$ so that the iteration can be rewritten as:
	\begin{equation}\label{eq:pwithsigma2}
		\begin{aligned}
			{\bp}_{j+1}& = {\bp}_j - \omega  \left(\sum_{k=0}^D\left(\sum_{s\in \varpi(k,j)} \hbS\Fs\right) \bS\Gm {\bp}_{j-k} + \hbb\right)
		\end{aligned}
	\end{equation}
	
	\subsubsection{Tools for convergence study}
	
	The asynchronous Richardson iteration was the object of~\cite{chow2021} in the case of a maximal delay of 2. In order to extend the method, we rely on the theory of paracontractions~\cite{paracontractions92}. 
	
	Let $(T_m)$ be a finite family of paracontractions with a common fixed point $\hat{x}$ in some Hilbert space $E$. In other words:
	\begin{itemize}
		\item $\forall x\in E,\ \|T_m(x)-\hat{x}\|<\|x-\hat{x}\|$ or $T_m(x)=x$,
		\item $\forall m,\ T_m(\hat{x})=\hat{x}$.
	\end{itemize} 
	Then a sequence of the form:
	\begin{equation}
		x_{j+1} = T_{m(j)}(x_j)
	\end{equation}
	converges to $\hat{x}$, assuming that all the paracontractions $(T_m)$ are sufficiently frequently activated.
	
	\subsubsection{Analysis of Global/Local coupling} 
	In order to make appear paracontraction, we assume a non-zero delay $D>0$, and we work in the ``history space'' obtained by concatenating the last $(D+1)$ values of ${\bp}_{j}$.
	
	We can rewrite the history at iteration $j+1$ as:
	\begin{equation}\label{eq:Bj}\begin{aligned}
			&	\begin{pmatrix}
				{\bp}_{j+1} \\ {\bp}_{j} \\ \vdots \\ {\bp}_{j-D+1}
			\end{pmatrix} = \underset{\bB_j}{\underbrace{\begin{pmatrix}
						\bI - \omega\bX_{j,0} & -  \omega\bX_{j,1} & \ldots & -  \omega\bX_{j,D}\\
						\bI & 0 & \ldots & 0 \\
						0 &\bI & 0 &\ldots\\
						\ldots & 0 &\bI &0
			\end{pmatrix}}}
			\begin{pmatrix}
				{\bp}_{j} \\ {\bp}_{j-1} \\ \vdots \\ {\bp}_{j-D}
			\end{pmatrix} - \begin{pmatrix}
				\omega\tilde{\bb} \\ 0 \\ \vdots \\ 0
			\end{pmatrix} \\
			&\text{with } \mathbf{X}_{j,k} =  \left(\sum_{s\in \varpi(k,j)} \hbS\Fs\right)\bS\Gm
		\end{aligned}
	\end{equation}
	Since $\forall j,\ \sum_k \bX_{j,k} \hbp + \tilde{\bb}=0$, the vector obtained by repeating the solution $\hbp$ of~\eqref{eq:solref} is a fixed point for the above iteration.
	
	In order to prove the paracontracting nature of the iteration, it suffices to prove that any matrix $\bB_j$ of~\eqref{eq:Bj} can be turned into contraction by correctly selecting the relaxation $\omega>0$. Since $\bB_j$ is a block companion matrix, it seems natural to study its spectrum and prove that it can be bounded by 1.  \medskip
	
	The eigenvalues $(\lambda)$ of $\bB_j$ are the roots of the polynomial:
	\begin{equation}\label{eq:roots0}
		\begin{aligned}
			\det\left((1-\lambda)\lambda^D\bI - \omega \sum_{k=0}^D \lambda^{D-k} \mathbf{X}_{j,k} \right)=0
		\end{aligned}
	\end{equation}
	This is the determinant of a real momic matrix polynomial~\cite{Gohberg2009}. 
	In order to benefit from the underlying symmetry, we can introduce the Cholesky factorization of $\bS\G=\bL\bL^T$, left-multiply the polynomial by $\bL^{-1}$ and right-multiply it by $\bL$, the roots of~\eqref{eq:roots0} are also the root of the polynomial $P_{j,\omega}(\lambda)$:
	\begin{equation}\label{eq:roots1}
		\begin{aligned}
			P_{j,\omega}(\lambda)&=\det\left((1-\lambda)\lambda^D\bI - \omega \sum_{k=0}^D \lambda^{D-k} \mathbf{\hat{X}}_{j,k} \right)=0
		\end{aligned}
	\end{equation}
	where $\mathbf{\hat{X}}_{j,k}= \bL^{-1}\mathbf{{X}}_{j,k}\bL =  \bL^{-1} \left(\sum_{s\in \varpi(k,j)} \hbS\Fs\right) \bL^{-T}$. %Note that the eigenvalues of $\mathbf{\hat{X}}_{j,k}$ are the eigenvalues of the generalized system $(\sum_{s\in \varpi(k,j)} \hbS\Fs,\bS\G)$.

	%Since the coefficient matrix are real, we have $P_{j,\omega}(\overline{\lambda})=\overline{P_{j,\omega}(\lambda)}$, so that the roots of the polynomial are complex conjugate.
	
	Using the absolute continuity of the roots of a polynomial with respect to its coefficients (see~\cite{hirose2020,parusinski:hal-01287852} for instance), we see that for a small enough $\omega$, the eigenvalues tend to concentrate around the roots of $P_{j,0}(\lambda)=\det((1-\lambda)\lambda^D\bI)$, that is to say around $0$ and  $1$.
	
\begin{figure}[ht]\centering
		\includegraphics[width=.6\textwidth]{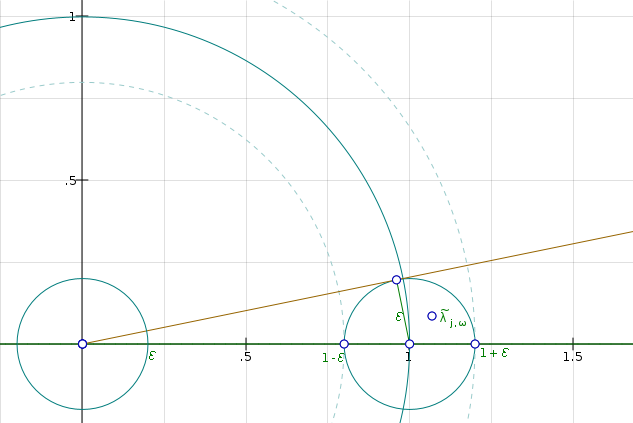}\caption{Constraining roots near $1$}\label{fig:lambdainC}
	\end{figure}

	Let $\tilde{\lambda}_{j,\omega}$ be one of the roots of $P_{j,\omega}$, and $\varepsilon = \min\left(\sin\left(\frac{\pi}{3D}\right),\frac{1}{2}\right)$, we can find $\omega_0$ such that $\omega<\omega_0$ $\Rightarrow$ $|\tilde{\lambda}_{j,\omega}-\tilde{\lambda}_{j,0}|<\varepsilon$. At that point, the roots that tend to zero have all modulus less than $\varepsilon<1$, only the roots that tend to 1 could pose a problem. In what follows, $\tilde{\lambda}_{j,\omega}$ is such a root that tends to 1, we can bound its modulus and argument, see Figure~\ref{fig:lambdainC}.
	
	$|\tilde{\lambda}_{j,\omega}-1|<\varepsilon$ implies that:
	\begin{equation}\label{eq:boundslam}
		\begin{aligned}
			&1-\varepsilon<|\tilde{\lambda}_{j,\omega}|<1+\varepsilon \\
			&	|\sin(\arg(\tilde{\lambda}_{j,\omega}))|< \varepsilon
		\end{aligned}
	\end{equation}
	For $\varepsilon=\sin{\frac{\pi}{3D}}$ and $0\leqslant k\leqslant D$, we have bounds on the modulus and on the reel part (symbol $\Re$):
	\begin{equation}\label{eq:boundslamk}
		\begin{aligned}
			(1-\varepsilon)^D&<|\tilde{\lambda}_{j,\omega}|^k<(1+\varepsilon)^D \\
			\Re(\tilde{\lambda}^k_{j,\omega})&=|\tilde{\lambda}_{j,\omega}|^k \cos(k\arg(\tilde{\lambda}_{j,\omega}))> \frac{(1-\varepsilon)^D}{2}
		\end{aligned}
	\end{equation}

	Let $\tbv_{j,\omega}$ be an eigenvector of the matrix polynomial associated with $\tilde{\lambda}_{j,\omega}$:
	\begin{equation}
		\begin{aligned}
			(1-\tilde{\lambda}_{j,\omega})\tilde{\lambda}_{j,\omega}^D\tbv_{j,\omega} - \omega \sum_{k=0}^D \tilde{\lambda}_{j,\omega}^{D-k} \mathbf{\hat{X}}_{j,k}\tbv_{j,\omega} &=0 
		\end{aligned}
	\end{equation}	
	We can left-multiply the expression by the Hermitian transpose $\tbv_{j,\omega}^H$: 
	\begin{equation}
		\begin{aligned}
			(1-\tilde{\lambda}_{j,\omega})\tilde{\lambda}_{j,\omega}^D \tbv^H_{j,\omega}\tbv_{j,\omega}- \omega \sum_{k=0}^D \tilde{\lambda}_{j,\omega}^{D-k} \tbv^H_{j,\omega}\mathbf{\hat{X}}_{j,k}\tbv_{j,\omega} &=0 \\
		\end{aligned}
	\end{equation}		
	To simplify, $\tbv_{j,\omega}$ can be chosen of unit Euclidean norm.  
	For $\omega<\omega_0$ we have $\tilde{\lambda}_{j,\omega}\neq 0$, and then:		
	\begin{equation}
		\begin{aligned}		
			\tilde{\lambda}_{j,\omega} &= 1-\omega \sum_{k=0}^D \frac{\|\tbv_{j,\omega}\|^2_{\mathbf{\hat{X}}_{j,k}}}{\tilde{\lambda}_{j,\omega}^{k}} \\
			|\tilde{\lambda}_{j,\omega}|^2 &= 1-	2\omega \sum_{k=0}^D \frac{\Re(\tilde{\lambda}_{j,\omega}^k)\|\tbv\|^2_{\mathbf{\hat{X}}_{j,k}}}{|\tilde{\lambda}_{j,\omega}|^{2k}} + \omega^2 \left|\sum_{k=0}^D \frac{\|\tbv\|^2_{\mathbf{\hat{X}}_{j,k}}}{\tilde{\lambda}_{j,\omega}^{k}}\right|^2
		\end{aligned}
	\end{equation}
	Using~\eqref{eq:boundslamk}, we have:
	\begin{equation}
		\begin{aligned}		
			|\tilde{\lambda}_{j,\omega}|^2 & < 1- \omega \frac{\left(\sum_{k=0}^D \|\tbv\|^2_{\mathbf{\hat{X}}_{j,k}}\right)}{(1+\varepsilon)^D}+ \omega^2 \frac{\left(\sum_{k=0}^D \|\tbv\|^2_{\mathbf{\hat{X}}_{j,k}}\right)^2}{(1-\varepsilon)^{2D}}
		\end{aligned}
	\end{equation}
	The sum of norms is simplified because each subdomain appears only once:
	\begin{equation}
		\begin{aligned}
			\sum_{k=0}^D \|\tbv\|^2_{\mathbf{\hat{X}}_{j,k}} & = \sum_{k=0}^D \tbv_{j,\omega}^H \mathbf{\hat{X}}_{j,k}\tbv_{j,\omega} = \tbv_{j,\omega}^H\bL^{-1}\left(\sum_{k=0}^D  \sum_{s\in \varpi(k,j)} \hbS\Fs \right)\bL^{-T}\tbv_{j,\omega} \\
			&= \tbv_{j,\omega}^H\bL^{-1} \left(\sum_{s=0}^N \hbS\Fs \right)\bL^{-T}\tbv_{j,\omega} 
		\end{aligned}
	\end{equation}
	Since  $\tbv_{j,\omega}$ is of unit Euclidean norm, the term above can directly be bounded by the extremal eigenvalues of $\bL^{-1} \left(\sum_{s=0}^N \hbS\Fs \right)\bL^{-T}$ which coincide to the generalized eigenvalues of the pair of matrices $(\sum_{s=0}^N \hbS\Fs,\bS\G)$:
	\begin{equation}
		\begin{aligned}
			\alpha_{\min} &\leqslant \sum_{k=0}^D \|\tbv\|^2_{\mathbf{\hat{X}}_{j,k}} \leqslant \alpha_{\max}\\
			\text{where the }(\alpha)\text{ solve } &\det\left(\left(\sum_{s=0}^N\hbS\Fs\right) + \alpha\bS\G\right)=0
		\end{aligned}
	\end{equation}
	
	We thus obtain the upper bound:
	\begin{equation}
		\begin{aligned}		
			|\tilde{\lambda}_{j,\omega}|^2 &\leqslant 1 - \omega \frac{\alpha_{\min}}{(1+\varepsilon)^D} + \omega^2  \frac{\alpha_{\max}^2}{(1-\varepsilon)^{2D}},\qquad \forall 0<\omega<\omega_0
		\end{aligned}
	\end{equation}
	This is a bound of the form $|\tilde{\lambda}_{j,\omega}|^2 \leqslant 1 - A \omega + B\omega^2$ (with $0<A<B$) which is a second degree polynomial in $\omega$, and which is less than 1 for $0<\omega<A/B$. As a consequence:
	\begin{equation}
		|\tilde{\lambda}_{j,\omega}|<1 \text{ for } 0<\omega< \omega_\mathrm{async}= \min \left(\omega_0, \frac{(1-\varepsilon)^D\alpha_{\min}}{(1+\varepsilon)^{2D}\, \alpha_{\max}^2 }\right)
	\end{equation}
	
	This is probably an extremely crude bound, but it has the advantage to only depend on $D$ and not on the configuration of the iteration (index $j$). Thus, such a relaxation makes any $\bB_j$ a paracontraction, and it makes the asynchronous iteration converge.
	
	\begin{remark}
		For the synchronous iteration, the bound can be derived from~\eqref{eq:roots0} with $D=0$, it is $0<\omega<\omega_\mathrm{sync} = \frac{2}{\alpha_{\max} }$. Note that $\omega_\mathrm{sync}>\omega_\mathrm{async}$.
	\end{remark}

	\subsection{Implementation details}\label{subsec:Implementations}
	\input{Implementations}

	%%%%%%%%%%%%%%%%%%%%%%%%%%%%%%%  APPLICATIONS %%%%%%%%%%%%%%%%%%%%%%%%%%%%%%%%%%%%%%%%%%%%
	\section{Applications}\label{sec:applications}
	\input{Applications_Lineaires}

	\section{Conclusion}
	An asynchronous version of the non-intrusive global/local computation method has been presented for linear elliptic problems, starting from the new interpretation of the method as a right-preconditioned primal domain decomposition method. A proof of convergence has been established for the discretized system using paracontractions techniques. An implementation with MPI RDMA parallelization has been set. The coupling has been tested on linear thermal and elasticity problems involving up to hundreds of patches. The performance in terms of computation time is convincing: the asynchronous method (with hand tuned relaxation) is faster than the synchronous solver with Aitken's dynamic relaxation on a cluster of heterogeneous machines.
	
	Future work should focus on finding an efficient estimation of the optimal relaxation for the asynchronous iteration.
	
\section*{Acknowledgments}
This work was partly funded by the French National Research Agency as part of project ADOM, under grant number ANR-18-CE46-0008.
\newpage

\bibliography{biblio}
\bibliographystyle{plain}

\end{document}

%% file: GLprincipLin.tex
The framework chosen to develop the method is the one of linear elliptic problems. This corresponds to certain thermal or elasticity static problems. We propose to derive the method as an evolution of the submodeling technique, we also give another (original) interpretation in terms of domain decomposition method.

\subsection{Principle of the method}
\begin{figure}[H]
	\null\hfill
	\begin{subfigure}[b]{0.3\linewidth}
		\centering
		\includegraphics[width=\textwidth]{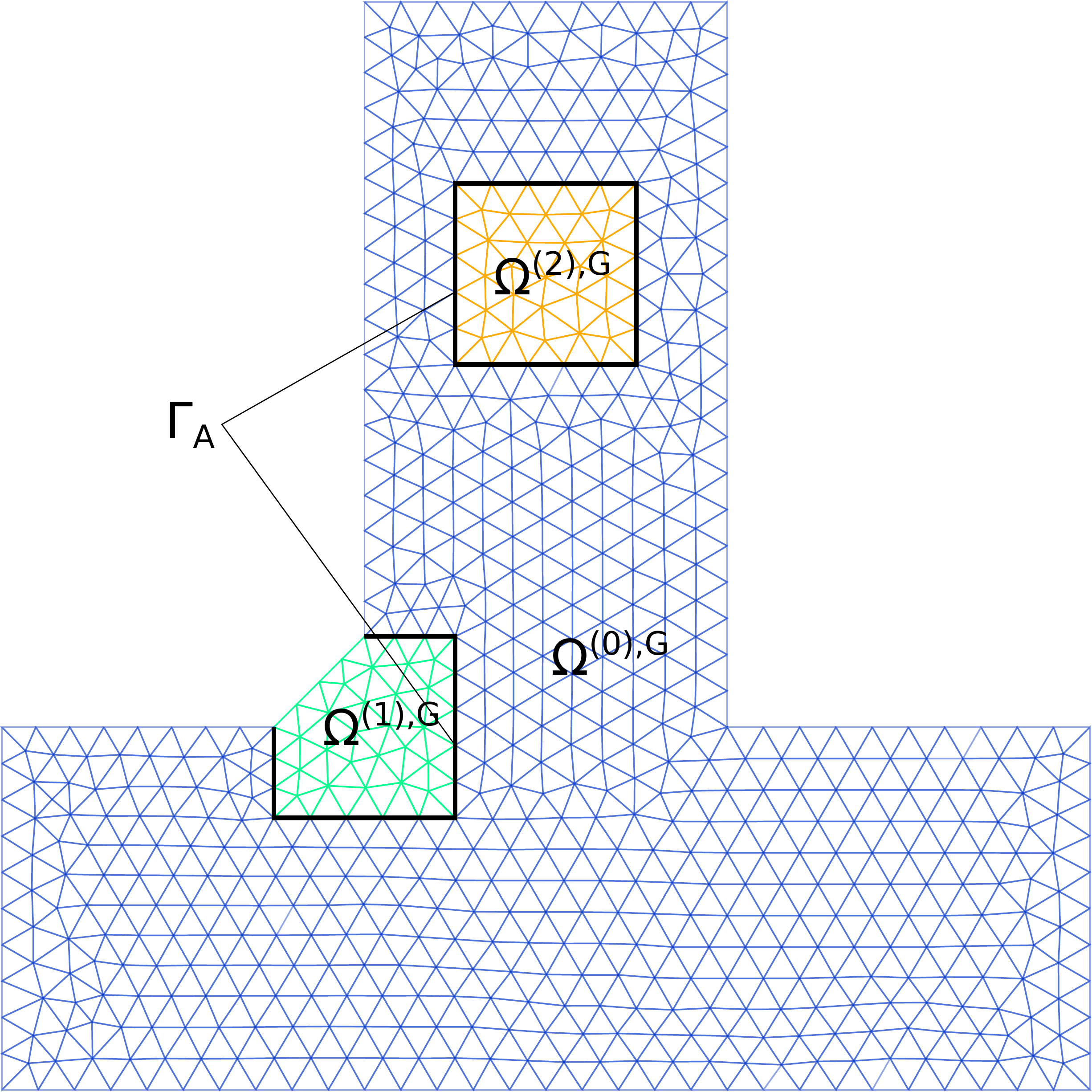}
		\caption{Global problem}\label{Global problem}
	\end{subfigure}
	\hfill
	\begin{subfigure}[b]{0.3\linewidth}
		\centering
		\includegraphics[width=.5\textwidth]{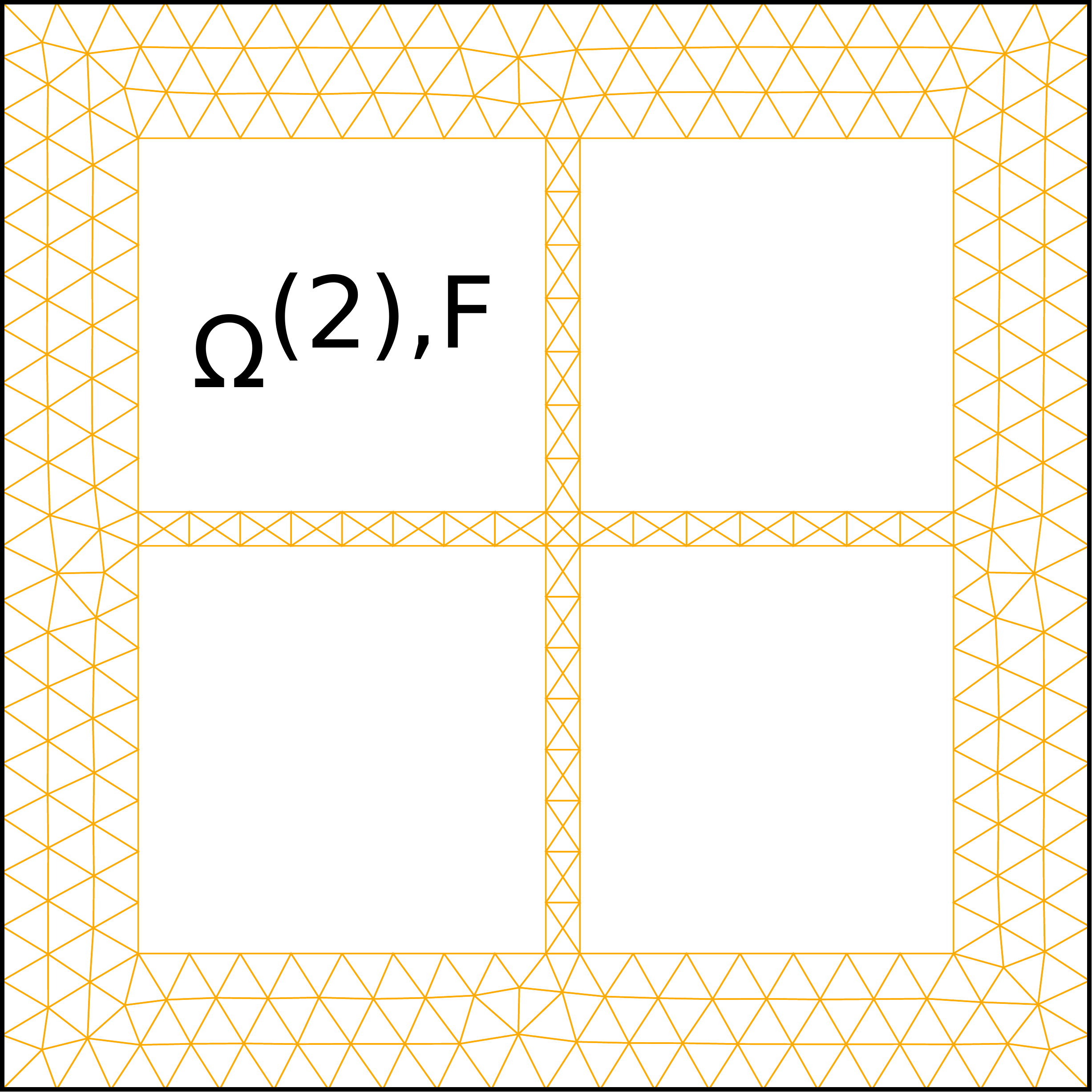}
		
		\null
		
		\includegraphics[width=.5\textwidth]{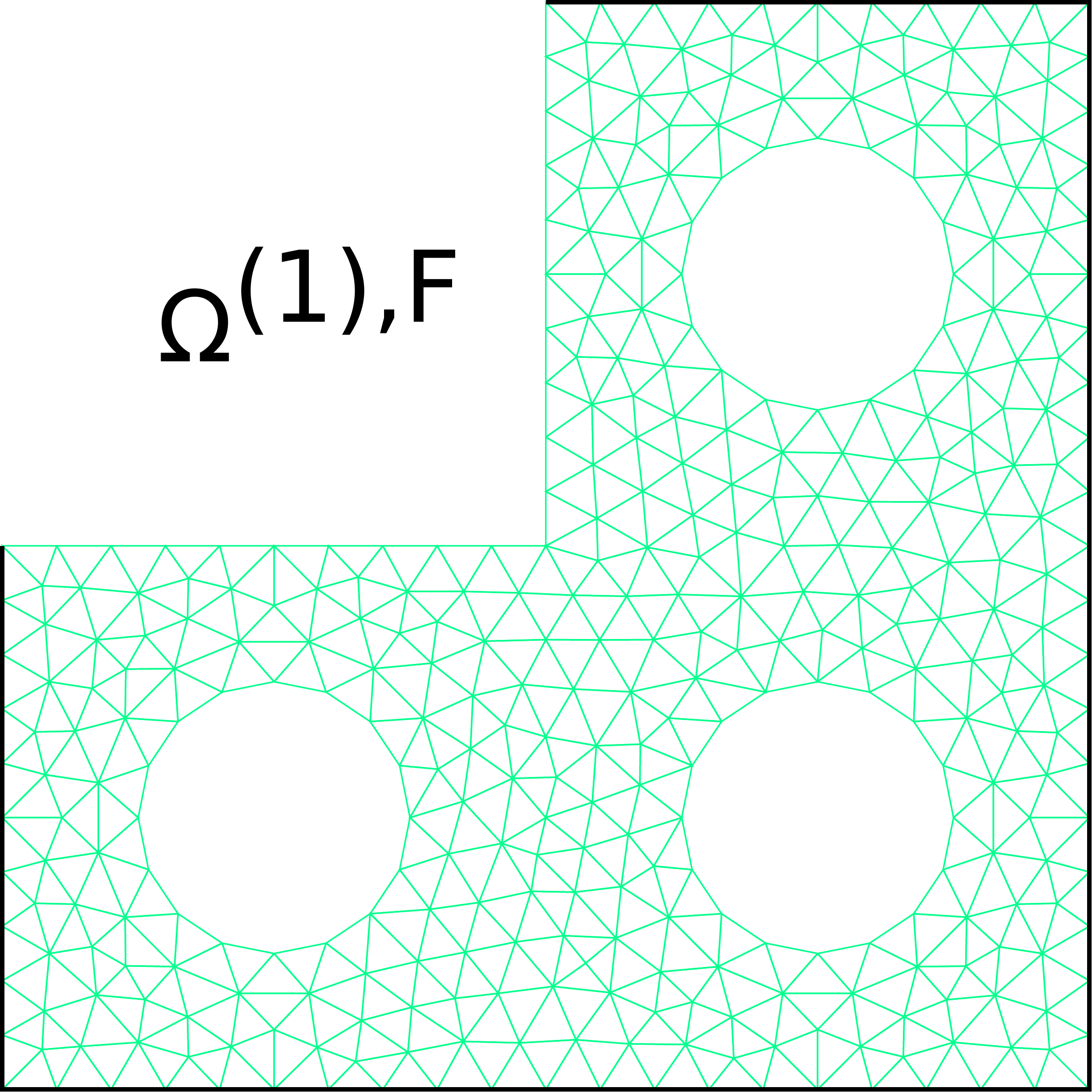}
		\caption{Refined zones of interest}	\label{Zones of interest}
	\end{subfigure}
	\begin{subfigure}[b]{0.3\linewidth}
		\centering
		\includegraphics[width=\textwidth]{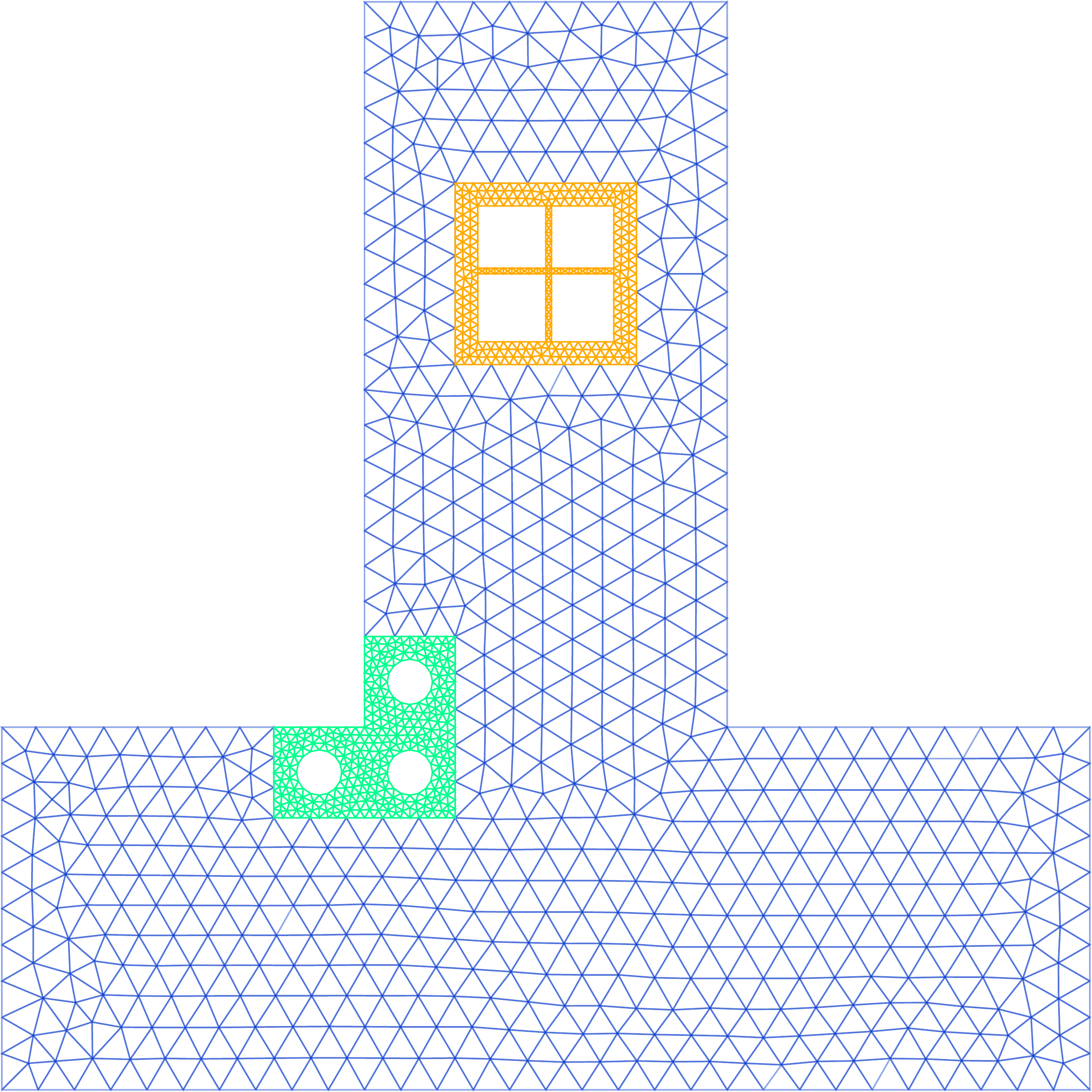}
		\caption{Reference problem}	\label{Reference problem}
	\end{subfigure}
	\hfill\null
	\caption{Models and subdomains for the Global/Local coupling}\label{fig:scenario}	
\end{figure}

\begin{figure}[H]
	\null\hfill
	\begin{subfigure}[b]{0.3\linewidth}
		\centering
		\includegraphics[width=\textwidth]{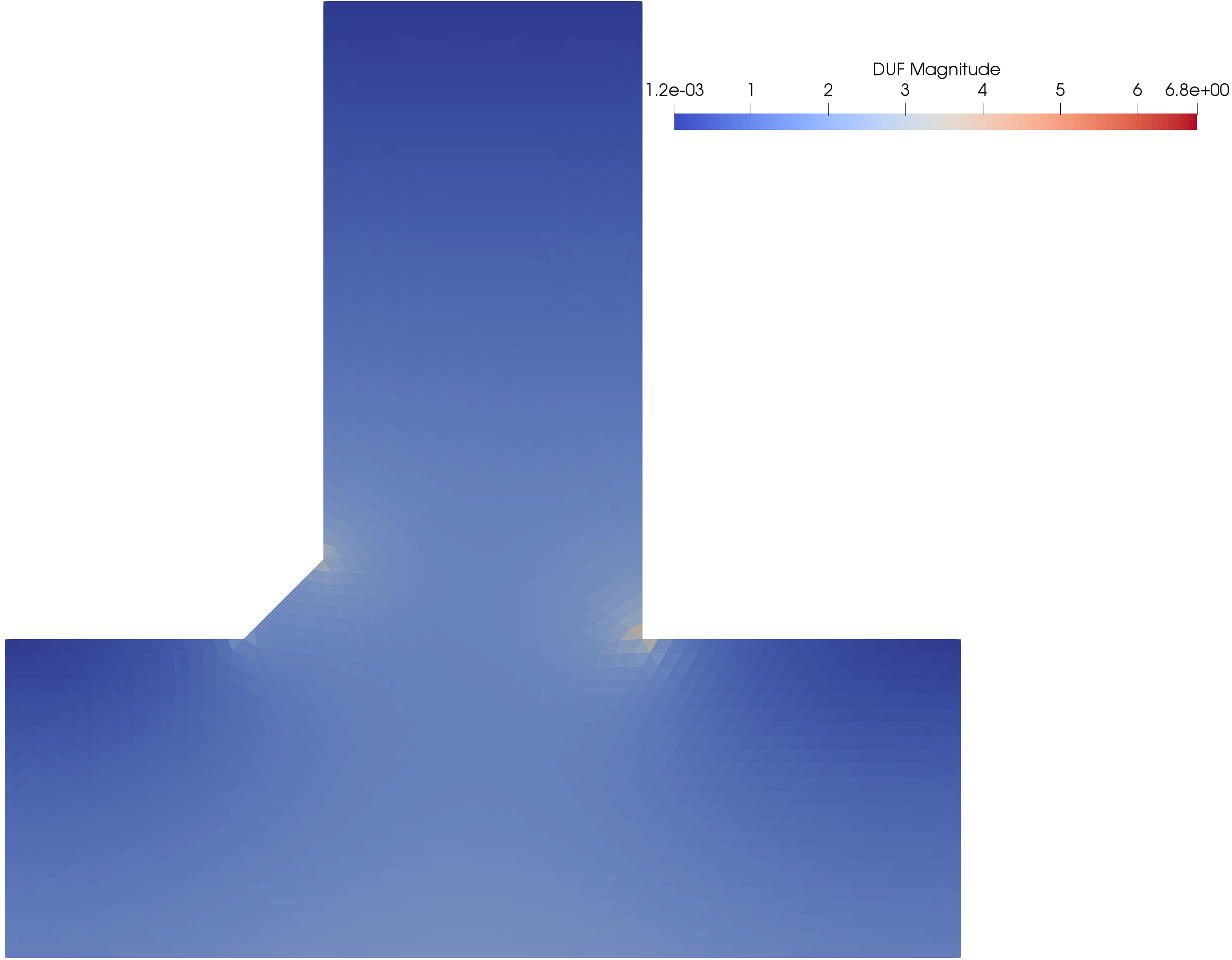}
		\caption{Global solution}\label{fig:scenario_vmglo}
	\end{subfigure}
	\hfill
	\begin{subfigure}[b]{0.3\linewidth}
		\centering
		\includegraphics[width=\textwidth]{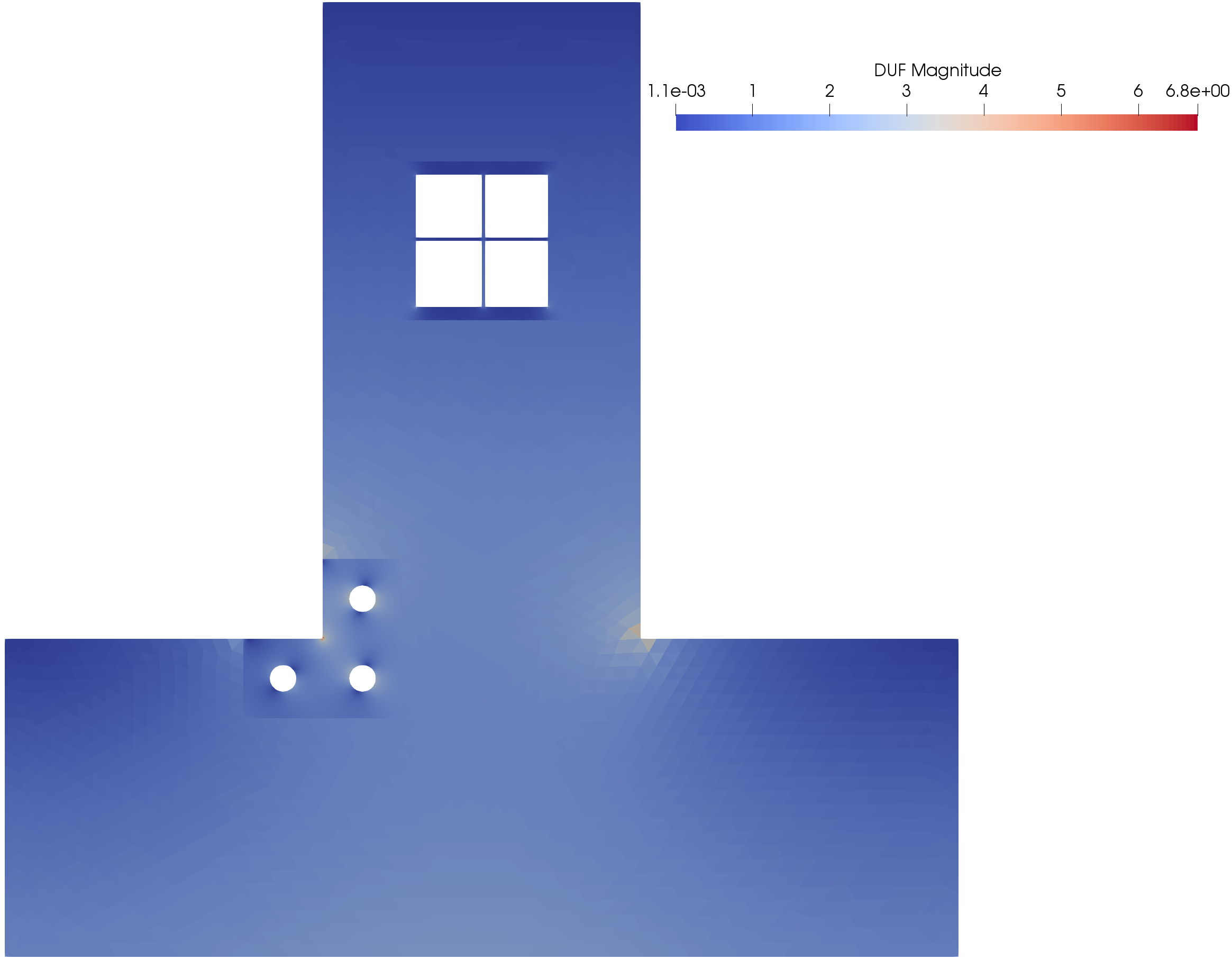}
		\caption{Submodeling solution}\label{fig:scenario_vmsub}
	\end{subfigure}
	\hfill
	\begin{subfigure}[b]{0.3\linewidth}
		\centering
		\includegraphics[width=\textwidth]{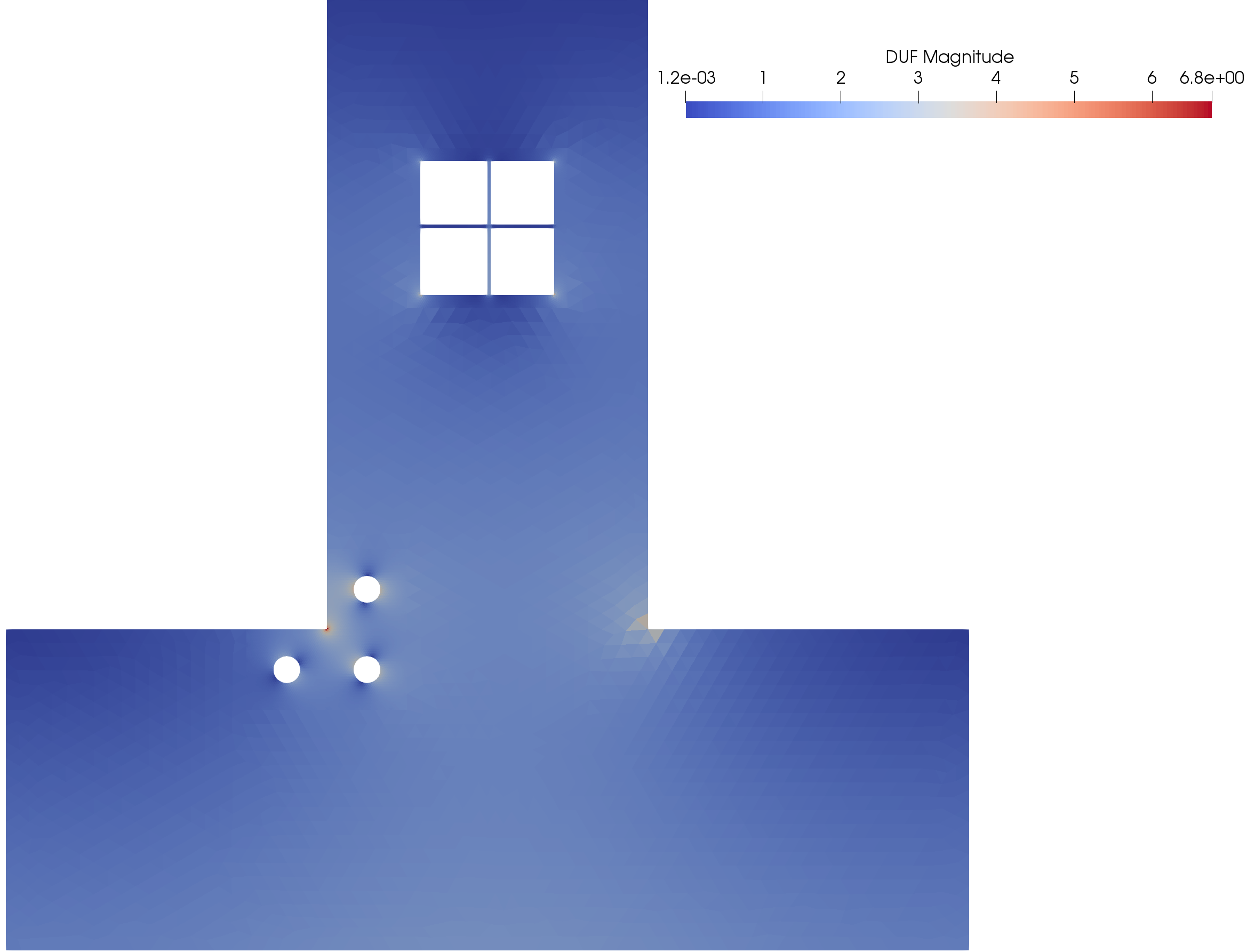}
		\caption{Reference solution}	\label{fig:scenario_vmref}
	\end{subfigure}
	\caption{Comparison of the norm of the heat flux for the submodeling and reference approaches (thermal problem)}\label{fig:scenariovm}	
\end{figure}

The classical scenario is illustrated on Figure~\ref{fig:scenario}. A linear Global coarse model is used to describe a large structure. After the initial computation (Figure~\ref{fig:scenario_vmglo}), some zones of interest $\Omega\Gs$  $(s>0)$ are selected because some criterion has been exceeded or because it was known from the beginning that some details were missing in the Global model. This is the case for our illustration where geometrical details and adapted meshes are introduced in the Fine modeling of the zones of interest $\Omega\Fs$. Material laws could also be modified by the introduction of some heterogeneity. Fine computations are run in parallel on the patches using the Global solution as Dirichlet boundary condition (for $s>0$, the interior of the Fine and Global subdomains may differ, but their interface $\Gamma\s$ must be the same $\Gamma\s=\Omega\cap\partial\Omega\Gs=\Omega\cap\partial\Omega\Fs$). 

This sequence of computations corresponds to the (in)famous submodeling technique which is known to result in large errors because the effects of Fine patches are not sent back to the Global model, and interactions between patches are thus impossible to be accounted for.

The error can be materialized by the lack of balance of the fluxes between the Global zone not covered by patches, denoted by $\Omega\z$ and the Fine models. As can be seen on Figure~\ref{fig:scenario_vmsub}, which shows the norm of the heat flux and where the Fine models overwrite the Global ones. There is a discontinuity at the interface which does not exist in the Reference computation where all interactions are taken into account; see Figure~\ref{fig:scenario_vmref} which corresponds to a direct computation of the Reference model where the zones of interest are described with the Fine models, see Figure~\ref{Reference problem}.

The Global/Local coupling is a simple iterative technique (a Richardson iteration for its simpler version) aiming at obtaining the Reference solution from computations carried on the Global and Fine models (that is to say without the potentially cumbersome creation of the Reference model) with minimal intervention on the models and software.

\subsection{Derivation of the Global/Local coupling}
There exist many ways to derive the Global/Local coupling. This subsection just sets up the method, the convergence of the asynchronous iteration being the subject of the next section.

We use boldface for discrete (nodal) quantities, lower case for vectors and upper case for matrices. 

\subsubsection{Global problem}

The Global problem is the classical finite element discretization of a coarse model of the structure, with one extra interface load. Let $\bp$ denote the vector of nodal fluxes applied on the interface nodes $\Gamma=\bigcup_{s=0}^N\Gamma\s$. To position the interface in the Global domain we introduce the boolean trace operator $\bT\G : \Omega\G\to\Gamma$, its transpose is the extension-by-0 operator.

The discrete Global problem can be written as:
\begin{equation}\label{eq:global}
	\left|	\begin{aligned}
		&\text{For given } \bp \text{ on }\Gamma,\text{ find }\bu\G \text{ in }\Omega\G, \text{ such that}\\
		&	\bK\G\bu\G = \f\G + \bT\GT \bp
	\end{aligned}\right.
\end{equation}
where one can recognize the symmetric definite positive stiffness matrix $\bK\G$, the vector of generalized loads $\f\G$, the vector of unknowns $\bu\G$.

The interface load is non-standard since it is a Neumann condition applied on an immersed surface. This corresponds to imposing a flux discontinuity in the Global model. It appears that such a load can easily be applied in industrial software, and the Global solution is obtained with a classical solver.
\medskip

In order to single out the contribution of subdomains, we introduce the boolean assembly operators $\bA\s:\ \Gamma\Gs\to\Gamma\G$ as classically encountered in the primal domain decomposition methods, see~\cite{GOSSELET.2007.1} for instance. Their transpose enables us to restrict some Global interface data to the boundary of a subdomain.

\subsubsection{Fine problems}
The Fine problems are set on the discretized subdomains $\Omega\Fs$. Boolean matrix $\bT\Fs$ is the trace operator on the Fine mesh $\Omega\Fs\to\Gamma\Fs$. For a good matching of the models, the interface is assumed to suit edges of the Fine elements. Anyhow, we do not require matching Global and Fine discretization, and we introduce Global-to-Fine transfer matrix $\bJ\s$ which enables us to define Fine Dirichlet problems with boundary conditions coming from the Global model.

The fine problems can be written as:
\begin{equation}\label{eq:sdF}
	\left|	\begin{aligned}	
		&\text{Given }\bu\G_\Gamma\text{ on }\Gamma; \forall s>0, \text{ find }\bu\Fs \text{ in }\Omega\Fs \text{ and }\lam\Fs \text{ on }\Gamma\s\text{ such that}\\
		&\bK\Fs\bu\Fs + \f\Fs = \bT\FsT\lam\Fs \\
		&\bT\Fs \bu\Fs = \bJ\s \bA\sT \bu\G_\Gamma\\
	\end{aligned}		\right.
\end{equation}

\subsubsection{Reference problem}
The Reference problem is the collection of Fine problems connected to the same interface displacement $\bu\G_{\Gamma}$ and such that the nodal reactions are in balance once projected back on the Global interface. 

First, we need to clarify the role played by Subdomain $0$, which might be non-existent. It is a subdomain, sometimes called Complement domain in the Global/Local literature, where the Fine and Global model coincide (same geometry $\Omega\z$, same properties, same load, same approximation). Its main role is to help process the nodal reaction $\lam\z$:
\begin{equation}\label{eq:sd0}
	\lam\z = \bT\z(\bK\z \bu\z - \f\z)
\end{equation}

We are now in position to formulate the Reference problem:
\begin{equation}\label{eq:reference}
	\left|	\begin{aligned}
		&	\text{Find }\bu\G_{\Gamma} \text{ on }\Gamma \text{ s.t }\\
		&	\br_\Gamma:=-\left(\bA\z\lam\z + \sum_{s=1}^N \bA\s \bJ\sT\lam\Fs\right)= 0\\
		&   \text{ where the reactions are obtained from}~\eqref{eq:sdF}\text{ and}~\eqref{eq:sd0}.
	\end{aligned}\right.
\end{equation}
%The minus sign is meant to make $\br$  to recover the classical definition of the residual.
%\medskip

\subsubsection{Condensed problems}
As usual with domain decomposition methods, the process is fully driven by the convergence of interface quantities.
For the analysis of the method, it is thus convenient to condense these previous problems at the interface. 

We then deduce from the system \ref{eq:sdF} the Dirichlet-to-Neumann operator for the Fine problems which can be written as:
\begin{equation}
		\lam\Fs = \bS\Fs \bu\Fs_\Gamma - \bb\Fs \\
\end{equation}		
With: 
\begin{equation*}
	\begin{aligned}	
		\begin{cases}
			\bS\Fs &= \bK\Fs_{\Gamma\Gamma} - \bK\Fs_{\Gamma i}\bK\Fsm_{ii}\bK\Fs_{i\Gamma}\\
			\bb\Fs &= \f\Fs_{\Gamma} - \bK\Fs_{\Gamma i}\bK\Fsm_{ii}\f\Fs_{i}
		\end{cases} 
	\end{aligned}
\end{equation*}
where $\bS\Fs$ is the well-known Schur complement and $\bb\Fs$ is the condensed right-hand side.

We use then the same notation for the condensation of Global subdomains, we can rewrite the Global problem \ref{eq:global} as:
\begin{equation}\label{eq:discreteglo}
	\underset{\bS\G}{\underbrace{\left(\sum_{s=0}^N \bA\s \bS\Gs\bA\sT\right)}} \bu\G_\Gamma = 	\underset{\bb\G}{\underbrace{ \left(\sum_{s=0}^N \bA\s \bb\Gs \right)}} + \bp
\end{equation}

The reference then ends up to being:
 \begin{equation*}\label{eq:ddschurnl}
 	\left|\begin{aligned}
 		&		\text{Find }\hbp \text{ such that}\\
 		&	\sum_{s=0}^N \bA\s \bJ\sT\left(\bS\Fs\bJ\s \bA\sT \left(
 		\bS\Gm \left(\hbp + \bb\G\right) 
 		\right) - \bb\Fs\right) = 0
 	\end{aligned}\right.
 \end{equation*}

In order to ease the reading, we introduce the notations:
%\begin{equation}
%\hbS\Fs = \bA\s \bJ\sT\bS\Fs\bJ\s \bA\sT,\qquad {\hbb} = \sum_{s=0}^N \bA\s \bJ\sT(\bS\Fs\bJ\s \bA\sT \bS\Gm \bb\G-\bb\Fs)
%\end{equation}
\begin{equation}
	\begin{aligned}	
		\begin{cases}
	\hbS\Fs &= \bA\s \bJ\sT\bS\Fs\bJ\s \bA\sT \\
	 {\hbb} &= \sum_{s=0}^N \bA\s \bJ\sT(\bS\Fs\bJ\s \bA\sT \bS\Gm \bb\G-\bb\Fs)
		\end{cases} 
	\end{aligned}
\end{equation}
so that the system to be solved can be written as:
\begin{equation}\label{eq:solref}
	\left(\sum_{s=0}^N \hbS\Fs \right)\bS\Gm \hbp + \hbb =0
\end{equation}

This system can be viewed as the primal domain decomposition formulation \cite{let91} of the Reference problem $\left(\sum_{s=0}^N \hbS\Fs \right)\bu\G_\Gamma =\left(\sum_{s=0}^N  \bA\s \bJ\sT \bb\Fs \right)$ right-preconditioned by the Global problem $\bu\G_\Gamma = \bS\Gm(\bb\G+\hbp)$. This preconditioner is of course much less scalable than the classical BDD strategy~\cite{man93} where local inverses of the Fine representation are used in conjunction with a much smaller coarse (global) problem. But this preconditioner provides a pertinent initialization $\bu\G_{\Gamma,0} = \bS\Gm \bb\G$ and it can be expected to introduce less irregularity at the interface, making it useless to add an enriched (spectral) coarse problem~\cite{spillane:2013:feti_geneo_ijnme}. Contrarily to the BDD approach where Krylov solver is mandatory (because the spectrum of the preconditioned operator is bounded from below by~1~\cite{kla01}), the Global/Local coupling supports stationary iteration. More, the right-preconditioning does not modify the nature of the residual of the system to be solved, allowing flexibility, and in our context, asynchronism.

\subsubsection{Global/Local coupling}
The aim of the coupling is to achieve~\eqref{eq:reference} using~(\ref{eq:global},\ref{eq:sdF},\ref{eq:sd0}). To do so, a simple modified Richardson iteration is used. Starting from $\bp=0$, we compute $\bu\G$ as in~\eqref{eq:global}, then we use $\bu\G_{\Gamma}$ as a Dirichlet condition to compute the Fine reactions $\lam\Fs$ using~\eqref{eq:sdF}\text{ and}~\eqref{eq:sd0}, finally the residual $\br_\Gamma$ is the lack of balance between the nodal reactions as in~\eqref{eq:reference}. If the residual is not small enough, the interface load is updated as $\bp=\bp+\omega\br_\Gamma$.
It can be proved that under the chosen hypothesis, there exist $\omega_{\max}>0$ such that the iteration converges for all $0<\omega<\omega_{\max}$. In practice, dynamic relaxation through Aitken's $\delta^2$ gives excellent performance.

Algorithm~\ref{alg:staSit} corresponds to applying a modified Richardson iteration to~\eqref{eq:solref}. The relaxation parameter is discussed in the next section as a particular case of the asynchronous iteration. In practice, it is recommended to use dynamic relaxation with Aitken's formula.

\begin{algorithm2e}[H]\caption{Synchronous stationary iterations}\label{alg:staSit}\DontPrintSemicolon
	Initialization $\bp=0$, $\omega$ sufficiently small\;
	\While{$\|\br\|$ is too large}{
		Resolution of the Global system~\eqref{eq:global} or~\eqref{eq:discreteglo}, $\bu\G_\Gamma = \bS\Gm (\bp + \bb\G)$\;
		\If{$\Omega\z$ exists }{Post-processing~\eqref{eq:sd0}, $\bq\z:=\lam\z = \bS\z \bu\Gz_\Gamma - \bb\Gz $}
		Global \textit{scatters} $\bA\sT\bu\G_\Gamma$ to subdmains $s>0$\;
		\For{$s>0$}{
			Patch \textit{receives} $\bA\sT\bu\G_\Gamma$\;
			Fine solution~\eqref{eq:sdF}, $\lam\Fs = \bS\Fs\bJ\s\bA\sT\bu\G_\Gamma-\bb\Fs$\;
			Patch \textit{sends} of $\bq\s:=\bJ\sT\lam\Fs$ to the Global
		}
		Global \textit{gathers} all $\bq\s$\;
		Global computes residual $\br=-\sum_s \bA\s\bq\s$\;
		Global updates $\bp = \bp + \omega\br$}
\end{algorithm2e}

%% file: Implementations.tex
Several approaches are available in the literature to implement asynchronous model. In \cite{Gbikpi-Benissan, Gbikpi-Benissan3}, an efficient library is proposed for asynchronous domain decomposition solvers, based on classical non-blocking two-sided communications. In \cite{Yamazaki.19, Glusa} the use of one-sided communications, also known as MPI-RDMA (Remote Direct Memory Access), is considered. The one-sided communication is meant to reduce management overhead. Note that the performance of the RDMA strongly depends on the MPI implementation and the network hardware.

The basic idea is that each rank exposes a so-called \textit{window} of its local memory and grants other ranks write or read access. Ranks, in this case, are no longer identified as sender or receiver but as \textit{origin} rank who initiates the operation and \textit{target} rank. The latter does not participate in the data exchange.

\begin{figure}[ht] 
	\begin{subfigure}{0.49\linewidth}
		\centering
		\includegraphics[width=0.9\textwidth]{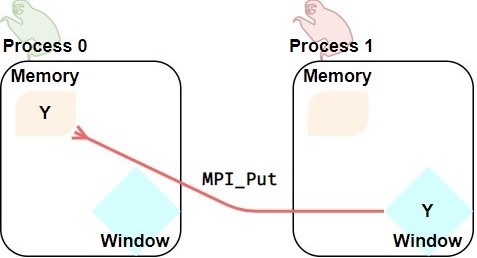}
		\caption{Put}
		\label{Put}
	\end{subfigure}
	\hfill
	\begin{subfigure}{0.49\linewidth}
		\centering
		\includegraphics[width=0.9\textwidth]{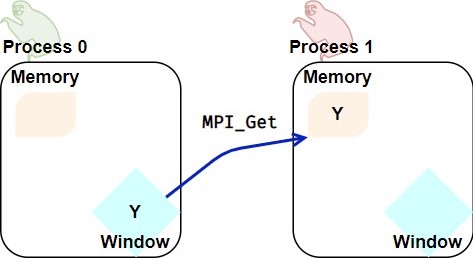}
		\caption{Get}
		\label{Get}
	\end{subfigure}
	\caption{One-sided communication concepts \cite{EuroCC}}
\end{figure}

The RMA-MPI workflow is based on the following five steps:
\begin{enumerate}
	\item \textbf{Allocation of the window} (local memory buffer accessible from other ranks).
	\item \textbf{Epoch opening}: beginning of the period when the window is open the other ranks.
	\item \textbf{Data accessing}:  
	Each origin rank can access the target ranks' window to \textit{Put} (write data) or to \textit{Get} (read data). See Figure~\ref{Put} where Processor 0 puts a data Y in Processor 1 window and Figure~\ref{Get} where Processor 1 gets a data Y from Processor 0 window.
	\item \textbf{Epoch closing}: the target rank which closes the windows ensures that all accesses are completed (local synchronization). At this point, the target rank can read and process the data put by other ranks.
	\item \textbf{Window freeing}: liberation of the memory buffer.
\end{enumerate}

To secure the data access in a window, one may consider two ways:
\begin{description}
	\item[Active synchronization] consists in performing a collective blocking call on both the \textit{target} and \textit{origin} using the \textit{MPI.Fence()} command at the beginning and at the end of the epoch to synchronize the data. 
	\item[Passive synchronization] emulates shared memory. The target processor is not involved in the management of the data, full asynchronous communication is possible. \textit{MPI.Lock(Target rank)} opens an epoch and allows the origin processor to access securely the target's window. The epoch is then closed by \textit{MPI.Unlock(Target rank)}. To ensure the completion of an operation within an epoch, one can use \textit{MPI.Flush(Target rank)}
\end{description}
Algorithm~\ref{alg:staAitRDMA} proposes an RDMA implementation of the asynchronous version of the Global/Local coupling algorithm \ref{alg:staAit} with passive synchronization. The principle is to have the subdomains compute whenever they idle and a new piece of information becomes available: a new interface Dirichlet condition for the \textit{fine} patches, any new interface nodal reaction for the \textit{global} model.

\begin{algorithm2e}[H]\caption{Asynchronous iterations using RDMA}\label{alg:staAitRDMA}\DontPrintSemicolon
	Window creation + Initialization $\bp=0$, $\omega$ sufficiently small \;
	MPI.Lock(target) (For all the window by specifying the specific target of each one)
	\While{$\|\br\|$ is too large}{
		\If{Rank 0 is available and detects at least one new $\bq\s$} {
			Resolution of the Global system~\eqref{eq:global} or~\eqref{eq:discreteglo}, $\bu\G_\Gamma = \bS\Gm (\bp + \bb\G)$\;
			\If{$\Omega\z$ exists }{Post-processing~\eqref{eq:sd0}, $\bq\z:=\lam\z = \bS\z \bu\Gz_\Gamma - \bb\Gz $}
			\textit{Put} $\bA\sT\bu\G_\Gamma$ in subdmains $s>0$ windows + \textit{Flush(subdomains s window)}\;
		}
		\For{$s>0$}{
			\If{Subdomain $s>0$ is available and detects new $(\bA\sT\bu\G)$}{
				Fine solution~\eqref{eq:sdF}, $\lam\Fs = \bS\Fs\bJ\s\bA\sT\bu\G_\Gamma-\bb\Fs$\;
				\textit{Put} of $\bq\s:=\bJ\sT\lam\Fs$ to the rank 0 window +
				\textit{Flush(0)}\;
			}
		}
		Global computes residual $\br=-\sum_s \bA\s\bq\s$\;
		Global updates $\bp = \bp + \omega\br$}
	MPI.UnLock(target) (For all the window by specifying the specific target of each one)
\end{algorithm2e}

\begin{remark}
An RDMA implementation of the synchronous coupling in algorithm~\ref{alg:staSit} is proposed  in Algorithm~\ref{alg:staSitRDMA}, it makes use of active synchronization with \textbf{MPI.Fence()}.
\end{remark}

\begin{algorithm2e}[H]\caption{Synchronous stationary iterations using RDMA}\label{alg:staSitRDMA}\DontPrintSemicolon
	Window creation + Initialization $\bp=0$, $\omega$ sufficiently small\;
	\While{$\|\br\|$ is too large}{
		MPI.Fence()(For the global displacement window) \;
		\If{rank == 0}{
			Resolution of the Global system~\eqref{eq:global} or~\eqref{eq:discreteglo}, $\bu\G_\Gamma = \bS\Gm (\bp + \bb\G)$\;
			\If{$\Omega\z$ exists }{Post-processing~\eqref{eq:sd0}, $\bq\z:=\lam\z = \bS\z \bu\Gz_\Gamma - \bb\Gz $}
			\textit{Put} $\bA\sT\bu\G_\Gamma$ in subdmains $s>0$\ windows ;}
		MPI.Fence()(For the global displacement window) \;
		MPI.Fence()(For the local nodal reaction) \;
		\If{rank != 0}{
			Fine solution~\eqref{eq:sdF}, $\lam\Fs = \bS\Fs\bJ\s\bA\sT\bu\G_\Gamma-\bb\Fs$\;
			Patch \textit{Put} $\bq\s:=\bJ\sT\lam\Fs$ in the rank 0 window
		}
		MPI.Fence()(For the local nodal reaction) \;
		MPI.Fence()(For the convergence detection window) \;
		\If{rank == 0}{
			Global computes residual $\br=-\sum_s \bA\s\bq\s$\;
			Global updates $\bp = \bp + \omega\br$}
		MPI.Fence()(For the convergence detection window) \;
	}
	Window free
\end{algorithm2e}

%This method permits to implement the synchronous solver where the data is in sync at the end of each iteration. This is generally implemented by the   command.% is generally which corresponds collective operation allowing to start an epoch at the beginning and closes it at the end.

%Both the \textbf{target} and \textbf{origin} have to call the method and the beginning and at the end of the epoch. Between the two calls, any amount of data access can be realized.

% \begin{itemize}
%	\item [$\bullet.$] \textbf{MPI\_Win\_flush(Target rank)}:  Completes the data transfer operations remotely at the target processor on the specified window, initiated by the origin processor calling the function.
%	\item [$\bullet.$] \textbf{MPI\_Win\_flush\_all()}:  Completes the data transfer operations remotely at all target processors on the specified window, initiated by the origin processor calling the function.
%	\item [$\bullet.$] \textbf{MPI\_Win\_flush\_local(Target rank)}: Completes the data transfer operations locally at the target processor on the specified window, initiated by the origin processor calling the function.
%	\item [$\bullet.$] \textbf{MPI\_Win\_flush\_local\_all()}: Completes the data transfer operations locally at all target processors on the specified window, initiated by the origin processor calling the function.
%\end{itemize}	

%% file: Applications_Lineaires.tex
%	To illustrate the theory presented above, we study linear and non-linear elliptic problems (\ref{linear_thermal}). The idea is to compare the performance in terms of computation time between the synchronous model and the asynchronous model through studies of low scalability, load imbalance, and finally, in an industrial case. 

%To illustrate the theory presented above, we have conducted studies on simple 2D academic cases with few patches and complex cases involving many patches.  This study allows us to see the effect of the load imbalance on the asynchronous and synchronous coupling and also weak scalability to see the behavior of both models by increasing the number of patches. Two problems have been studied; the first one corresponds to a linear thermic scalar problem, and the second one is a linear elasticity vector problem. 

To illustrate the theory presented above, we consider two kind of equations.
First the Poisson equation, which models thermal problems:
\begin{equation}\label{linear_thermal}
	\begin{aligned}
		\text{Find }u&:\Omega\subset\mathbb{R}^d\to\mathbb{R}\\
		\operatorname{div} (a \operatorname{grad}( u)) &= 1  \text{ in }\Omega\\
		u &= 0 \text{ on }\partial_d\Omega\\
		\frac{\partial u}{\partial n} &= 0 \text{ on }\partial\Omega\setminus\partial_d\Omega\\
	\end{aligned}
\end{equation}
for simplicity, we used unit source term and homogeneous boundary conditions. In some cases a contrast of conductivity coefficient $a$ is used.

Second, the linear elasticity equation:
\begin{equation}
	\begin{aligned}
		\text{Find }u&:\Omega\subset\mathbb{R}^d\to\mathbb{R}^d\\
		\operatorname{div} (\sigma ) + f &= 0  \text{ in }\Omega\\
		u &= 0 \text{ on }\partial_d\Omega\\
		\sigma \cdot n &= 0 \text{ on }\partial\Omega\setminus\partial_d\Omega\\
		\sigma &= \frac{E}{1+\nu}\left(\varepsilon(u)+\frac{\nu}{1-2\nu}\operatorname{tr}(\varepsilon(u)) I\right) \\ 
		\varepsilon(u) &=\frac{1}{2}  (\nabla u +  (\nabla u)^{T})\\
	\end{aligned}
\end{equation}
$E$ is Young's modulus, and $\nu=0.3$ is Poisson's coefficient. In some cases, a contrast of Young's modulus is used. The value of the source term f varies with the study cases.

We propose to assess the asynchronous Global/Local coupling on two academic examples: the simple 2D case of Figures~\ref{Global problem} and~\ref{Zones of interest}, and a more challenging 3D case involving many patches. In order to evaluate the performance we compare the following approaches:
\begin{itemize}
	\item non-relaxed synchronous iteration ($\omega=1$),
	\item Aitken-accelerated (synchronous) iteration,
	\item non-relaxed asynchronous iteration ($\omega=1$),
	\item asynchronous iteration with optimized relaxation.
\end{itemize}
Aitken's acceleration can be viewed as an efficient way to find a good dynamic relaxation. The optimized relaxation coefficient for the asynchronous iteration is obtained by trial-and-error. \medskip

Our Ethernet network does not support RDMA communication by default. It generates implicit synchronizations when we use \textit{MPI.Lock()} and \textit{MPI.Unlock()} commands to check if new data is available in the target processor. In order to achieve the best possible time, we used a computational sequence slightly different than Figure \ref{Asynchronous with wait}: processors always compute with the available data without checking for their novelty (thus possibly redoing the same calculus several times but never triggering unwanted sync).

Our code is realized in Python with \textit{mpi4py} module \cite{mpi4py}. It uses several other tools and software like \textit{GMSH} \cite{GMSH} to generate the geometries and meshes of the studied cases. For the finite element approximation, we use the \textit{GetFEM} library \cite{GetFEM}. 

The study was carried out with the cluster of the LMPS simulation center using several workstations with an Ethernet network. These machines are quite heterogeneous with 4 different generation of CPUs :(Intel(R) Xeon(R) CPU E5-1660 v3 (Haswell) @ 3.00GHz,  Intel(R) Xeon(R) CPU E5-2630 v4 (Broadwell) @ 2.20GHz, Intel(R) Xeon(R) Silver 4116 CPU (Skylake) @ 2.10GHz, Intel(R) Xeon(R) W-2255 CPU (Cascade Lake) @ 3.70GHz.

\medskip

\subsection{Simple 2D test-case}
To begin with the illustrations, we use the test-case of Figures~\ref{Global problem} and~\ref{Zones of interest} where the patches only introduce geometric alterations. The patches and the global model are treated on three different CPUs. 

As shown in Table~\ref{Mesh 2D}, the problem is of very small dimension, and the patches are well-balanced, which is in favor of synchronous algorithms.

\begin{table}[H]
	\centering
	%\small\addtolength{\tabcolsep}{-3pt}
	\begin{tabular}{|l|c|c|c|}
		\hline
		Problem&  \textbf{Global} & \textbf{$1^{st}$ zone of interest} & \textbf{$2^{nd}$ zone of interest}\\ \hline
		\#nodes &  701 &  381 &   379\\ \hline
	\end{tabular}
	\caption{Size of the domains or the 2D test case.}
	\label{Mesh 2D}
\end{table}

Tables~\ref{Meshes2Dth} and~\ref{Elasticity_2D} present the performance in terms of time and number of iterations. In the asynchronous cases, the number of local solves may differ from the number of iterations, so the range of the number of local solves is also indicated. 
For these small cases, Aitken remains unbeatable. We observe the interest of finding a good relaxation for the asynchronous iteration to perform better than the non-relaxed synchronous iteration. 

\begin{table}[H]
		\centering
		\small\addtolength{\tabcolsep}{-4pt}
		\begin{tabular}{|l|C{2.1cm}|C{2.1cm}|C{2.1cm}|C{2.1cm}|}
			\hline
			Variant&  \textbf{Sync.}\newline $\omega=1$ & \textbf{Sync.}\newline  Aitken & \textbf{Async.}\newline  $\omega=1$ &  \textbf{Async.}\newline  $\omega_\mathrm{opt}$ \\\hline
			{Time (s)}   &  0.22   & 0.12      & 0.3  & 0.19 \\ 
			\#iter. glob.  &  23   & 12     &  45 & 29  \\ 
			\#loc. sol. [min, max] &  $\cdot$   &  $\cdot$    &  [96,97] & [64,65]  \\ \hline
		\end{tabular}
		\caption{2D test-case: performance for thermal problem.}
		\label{Meshes2Dth}
\end{table}

\begin{table}[H]
	\centering
	\small\addtolength{\tabcolsep}{-4pt}
	\begin{tabular}{|l|C{2.1cm}|C{2.1cm}|C{2.1cm}|C{2.1cm}|}
		\hline
			Variant&  \textbf{Sync.}\newline $\omega=1$ & \textbf{Sync.}\newline  Aitken & \textbf{Async.}\newline  $\omega=1$ &  \textbf{Async.}\newline  $\omega_\mathrm{opt}$ \\\hline
		{Time (s)}   &  0.67  & 0.3      & 0.6  & 0.52 \\ 
		\#iter. glob. &  43   & 16     &  53  & 48  \\ 
		\#loc. sol. [min, max] &  $\cdot$   & $\cdot$     & [112,119]  & [100,107] \\ \hline
	\end{tabular}
	\caption{2D test-case: performance for the elasticity problem.}
	\label{Elasticity_2D}
\end{table}

What is more interesting to observe is the large amount of computation that can be done by the asynchronous solver thanks to the removal of waiting time. 

%{\color{red} on est bien dans un cas NO WAIT?? si oui il faut expliquer}

\subsection{Weak scalability 3D test-case}
Weak scalability tests aim at proving the ability of the method to solve large problems in reasonable time. 

\begin{figure}[H]
		\begin{subfigure}{.48\textwidth}
			\centering
			\includegraphics[width=0.9\textwidth]{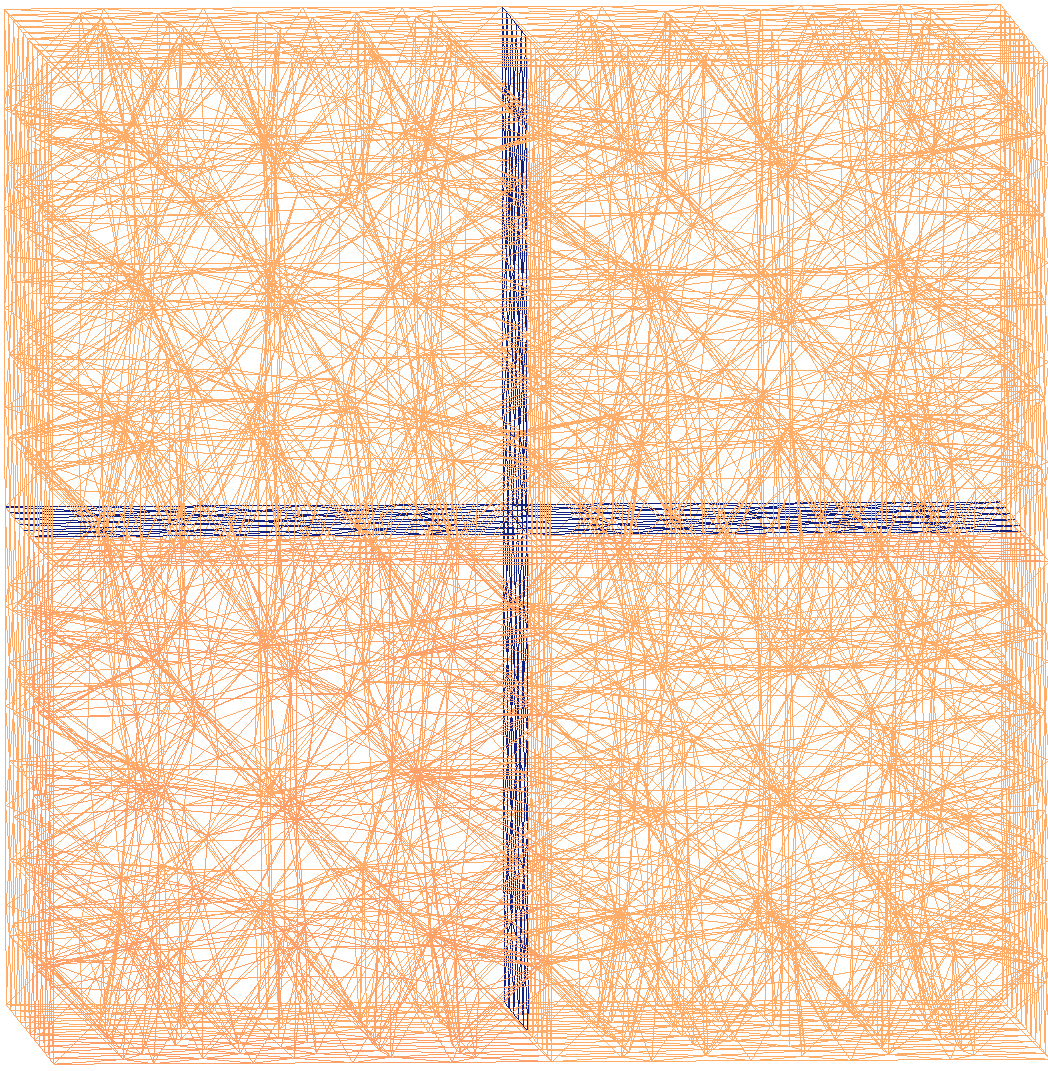}
			\caption{Global model}
			\label{Coarserepres}
		\end{subfigure}
		\hfill
		\begin{subfigure}{.48\textwidth}
			\centering
			\includegraphics[width=0.9\textwidth]{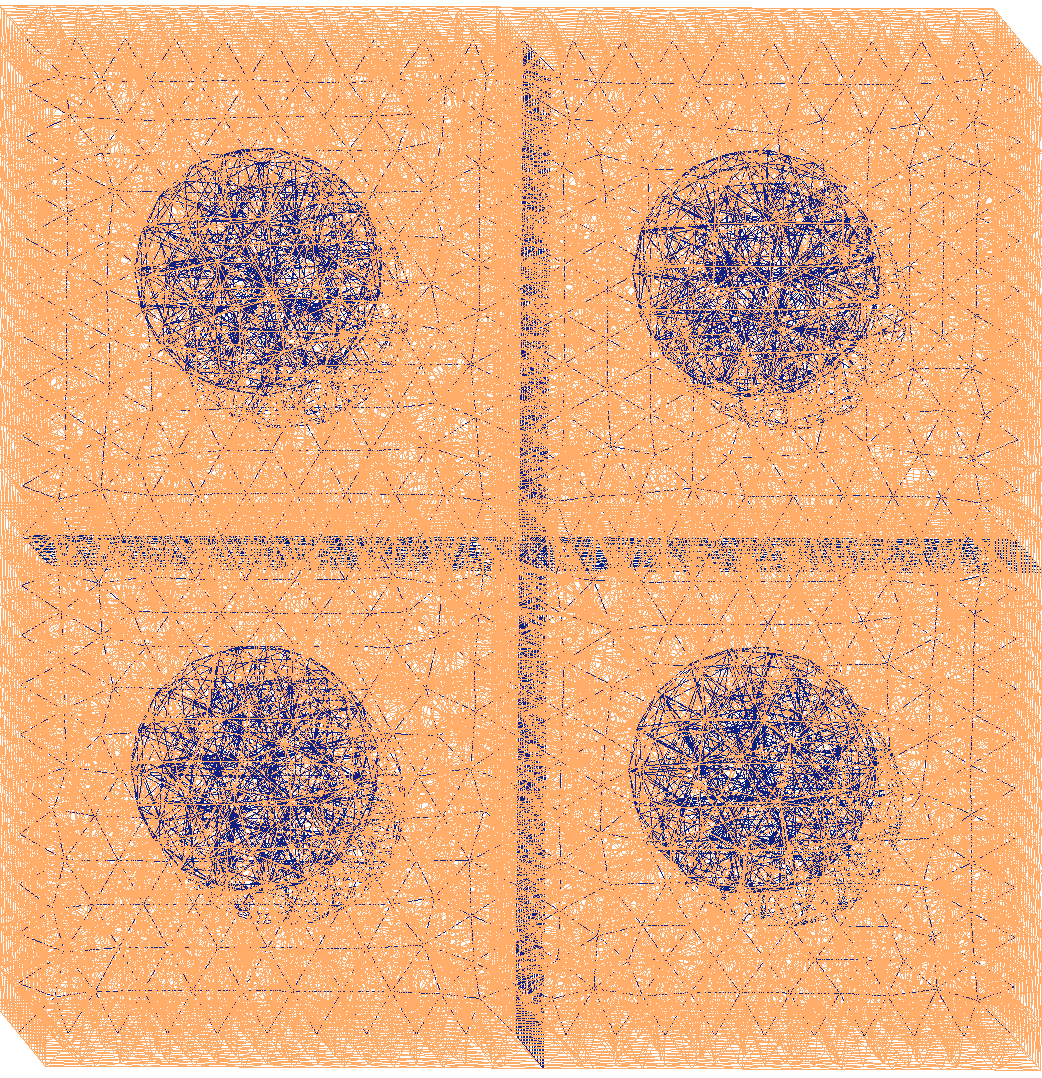}	
			\caption{Reference model}
			\label{Finerepres}
		\end{subfigure}
		\caption{Weak scalability test-case: $2\times 2\times 2$ subdomains}
	\end{figure}

In order to be able to generate test-cases with many patches, we created a cuboid geometry made out of $n^3$ ($n=2..7$) cube patches. As classically done for weak scalability assessment of domain decomposition methods, the size of the domain increases with the number of subdomains. Note that the whole domain is covered with patches ($\Omega\z=\emptyset$). The Global model is homogeneous, whereas the Local models contain one softer spherical inclusion, see Figures~\ref{Coarserepres} and~\ref{Finerepres}. One side of the Global model is submitted to Dirichlet conditions.

In the case of thermal problems, the inclusions have a diffusion coefficient 10 times lower than the rest of the domain, whereas in the elasticity case the Young's modulus in the inclusions is 100 times lower than in the rest of the domain.

%		\begin{figure}[ht]
%		\centering
%		\includegraphics[width=0.48\textwidth]{./Global_8_bis.png}	
%		\caption{8 subdomains with $\Omega^{0} = \emptyset$ }
%		\label{8_subd}
%	\end{figure}
%	Figure \ref{8_subd} shows the case with eight subdomains without complementary domain, in this case the structure is cover of patches. 
	
%	The subdomain's refined version has a heterogeneous spherical inclusion compared to the rest of the matrix.
% is ten times smaller in the sphere for linear thermal, while the coarse version is for both homogeneously made only of the matrix.
	
%\caption{3D domain decomposition with $\Omega^{0} = \emptyset$}
	
%	Figure \ref{Coarserepres}, shows each subdomain's coarse representation, which we refine in figure \ref{Finerepres} and add the spherical inclusion.
	
%	The idea of the study as presented above is to increase the number of subdomains while keeping the same cubic geometry, so we consider the cases with $2^{n}$ subdomains with $n = 2, ..., 7$.
	
Even if their meshes are not identical, the patches are well-balanced in terms of degrees of freedom and numerical complexity (since the problem is linear). Of course, the Global model grows along the study, from 8 times smaller than one patch to 3.7 times larger. This is a strong limitation of the method in comparison with classical domain decomposition methods were the coarse problem's growth is much more moderate. Table~\ref{Meshes} sums up the number of nodes for each case.
	
	\begin{table}[H]
		\centering
		\begin{tabular}{|l|c|c|c|c|c|c|}
			\hline
			\#subdomains &  \textbf{8} & \textbf{27} & \textbf{64} &  \textbf{125} &  \textbf{216}&  \textbf{343}\\ \hline
			\hline
			\#nodes of glob. problem   & 233 &   667 & 1449  & 2681  & 4465    &  6903   \\
			\#nodes of per loc. problem (= 1 subdomain) & 1858 &1858     & 1858      & 1858  & 1858  & 1858  \\ \hline
		\end{tabular}
		\caption{Number of nodes in the meshes for the weak scalability study.}
		\label{Meshes}
	\end{table}
	
%	For all the study, we keep the same value for our right-hand side : $f = 1$. Also, we keep the same Dirichlet condition.  $u = 0\text{ on }\partial_d\Omega$ corresponding to one side of the global cube.

%\begin{figure}[ht]
%	\begin{subfigure}{.49\textwidth}
%		\centering
%\includegraphics[width=0.99\textwidth]{./Thermal_weak_extensibility_bis.png}
%\caption{Thermal problem}
%\label{scalability_linear_thermal}
%	\end{subfigure}
%	\begin{subfigure}{.49\textwidth}
%		\centering
%\includegraphics[width=0.99\textwidth]{./Elasticity_weak_extensibility.png}
%\caption{Elasticity problem}
%\label{scalability_linear_elasticity}
%			\end{subfigure}
%		\caption{Time performance in the weak scalability study\color{red} peut-être trop petit}\label{fig:WStime}
%\end{figure}

\begin{figure}[H]
%\begin{subfigure}{.49\textwidth}
	\centering
	\includegraphics[width=0.8\textwidth]{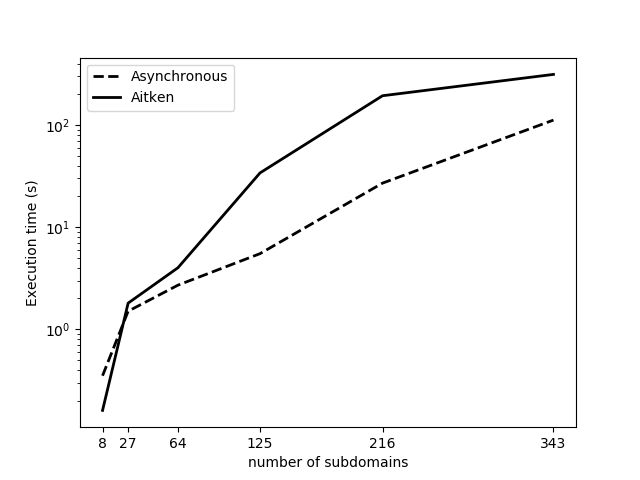}
	%\caption{Thermal problem}
	\caption{Time performance in the weak scalability study for linear thermal problem}
	\label{scalability_linear_thermal}
\end{figure}
\begin{figure}[H]
%\begin{subfigure}{.49\textwidth}
	\centering
	\includegraphics[width=0.8\textwidth]{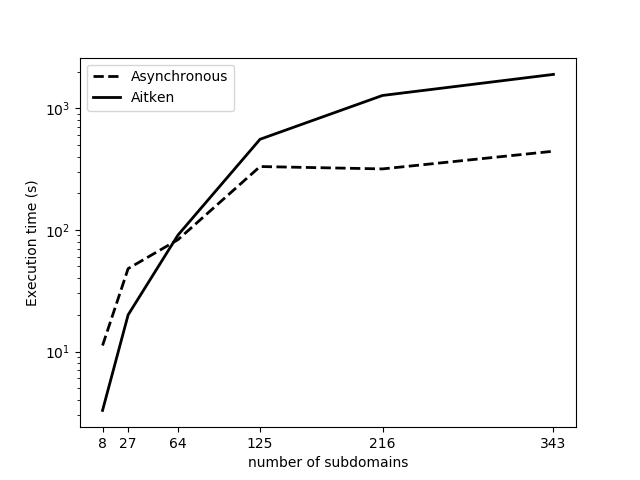}
	%\caption{Elasticity problem}
	\caption{Time performance in the weak scalability study for linear elasticity problem}
	\label{scalability_linear_elasticity}
\end{figure}

Figures~\ref{scalability_linear_elasticity} and \ref{scalability_linear_thermal} compare the performance in wall-clock time of the relaxed asynchronous iteration (with hand-tuned relaxation) and the synchronous iteration with Aitken's dynamic relaxation. We observe the good performance of the asynchronous version despite the good load-balancing.
	
For the small test-cases (8 and 27 subdomains), the size of the global problem is negligible compared to the size of the local problems. This means that the sequential phase of the synchronous coupling is realized very quickly and this leads to the Aitken accelerator being faster than the asynchronous solver. However, for 64 subdomains and more, this step becomes heavier and takes more synchronous time. For the asynchronous method, the Global solve is realized simultaneously with the local solves. Thus, the execution time increases very slightly from one case to another and remains 2 to 3 times less than for Aitken.

	\begin{table}[H]
		\centering
		\small\addtolength{\tabcolsep}{-4pt}
		\begin{tabular}{|l|c|c|c|c|c|c|}
			\hline
			\#patches &  \textbf{8} & \textbf{27} & \textbf{64} &  \textbf{125} &  \textbf{216}&  \textbf{343}\\ \hline \hline
			
			Aitken  \#iter.&  11  & 13    & 12 & 11& 11 & 11  \\\hline
%			Async. \#it. & 255 [32,39]   & 256 [43,74]     &  87 [49,153]  & 65 [84,207] & 69 [276,694] & 71 [407,2902]  \\
			Async. \#iter. glob. & 255    & 256      &  87   & 65 & 69  & 71   \\
			Async. \#loc. sol. [min, max] &  [32,39]   &  [43,74]     &   [49,153]  &  [84,207] &  [276,694] &  [407,2902]  \\ \hline
		\end{tabular}
		\caption{Weak scalability: Number of iterations in the thermal case.}
		\label{scalability_linear_thermal_table}
	\end{table}

	\begin{table}[H]
	\centering
	\small\addtolength{\tabcolsep}{-4pt}
	\begin{tabular}{|l|c|c|c|c|c|c|}
		\hline
		\#patches&  \textbf{8} & \textbf{27} & \textbf{64} &  \textbf{125} &  \textbf{216}&  \textbf{343}\\ \hline \hline
		
		Aitken \#iter.  &  22 &  21    & 25 & 25 & 26 & 29  \\\hline
%		Asynchronous &  2065[78 - 240]   & 1349[102 - 237]     &  372[128 - 475]  & 296[157 - 517] & 295[147 - 514] & 209[175 - 407]  \\ \hline
		Async. \#iter. glob. &  2065  & 1349     &  372  & 296 & 295 & 209 \\ 
		Async. \#loc. sol. [min, max] &  [78,240]   & [102,237]     &  [128,475]  & [157,517] & [147,514] & [175,407]  \\ \hline
	\end{tabular}
		\caption{Weak scalability: Number of iterations in the elasticity case.}
	\label{scalability_linear_elasticityl_table}
\end{table}

	Tables~\ref{scalability_linear_thermal_table} and~ \ref{scalability_linear_elasticityl_table} gather the number of iterations for each case. In the asynchronous case, the number of iterations (or Global solves) is given as well as the minimum and maximum numbers of patches' solve. 
	We see that the number of iterations barely varies in the synchronous experiments (in particular for the thermal problem) for all studied cases. 
	
	For the asynchronous solver, it can be seen that in the 8 and 27 patches cases where the global problem is very small, many more solves are performed by the global domain than by the local patches. Because of the non-waiting asynchronous model the global problem repeats several times the same calculation without having new information from the locals, however when the size of this problem increases (more than 64 subdomains), we begin to see that the patches make more repeated iterations while waiting for the update of the global problem which performs only a few iterations.
	
	Note the performance achieved in the elasticity case (2 times faster) despite the tremendous number of iterations (7 times more).

	\subsection{Poor load balancing}
\begin{figure}[H]
			\centering
			\includegraphics[width=0.68\textwidth]{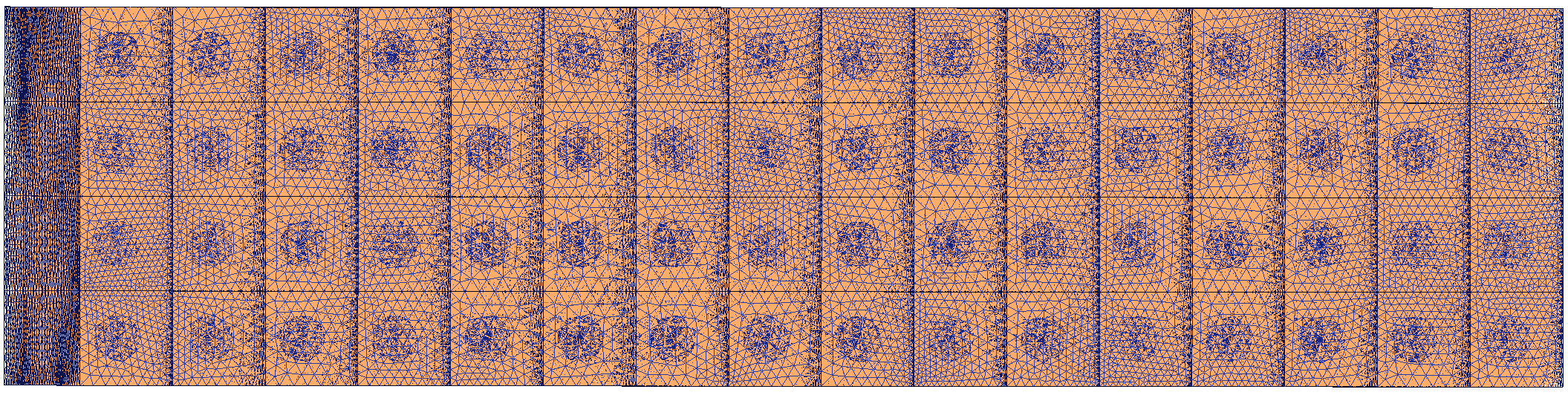}	
			\caption{Fine representation with unbalanced subdomains}
			\label{Fine}
	\end{figure}

We wish to evaluate the influence of a significant disequilibrium in the number of nodes to be handled by processors. We start from a geometry formed with a $16\times4\times4$  repetition of cubes with spherical inclusion (this time 1000 times stiffer than the rest of the domain), see Figure~\ref{Fine}.
%	\begin{figure}[t]
%		\centering
%		\includegraphics[width=0.68\textwidth]{./256.png}
%		\caption{256 subdomains with $\Omega^{0} = \emptyset$ }
%		\label{256}
%	\end{figure} 
%	Figure \ref{256} corresponds to the study case consisting of 256 subdomains.
%	
%	\begin{figure}[H]
%			\centering
%			\includegraphics[width=0.68\textwidth]{./256_deseq_Global.png}
%			\caption{Coarse representation}
%			\label{Coarse}
%	\end{figure}
%
%	Figure \ref{Coarse} corresponds to the rough representation of each of the subdomains. The assembly of these representations is thus what constitutes the global problem on the structure.
Each Fine subdomain has a randomly chosen number of nodes compared to the other subdomains, allowing to have very refined subdomains and others slightly refined. Table \ref{Meshes2} summarizes the number of nodes for the global problem and the smallest and largest number of nodes among the 256 Fine subdomains. We can see that the most refined subdomain is ten times larger than the least refined.
	
	\begin{table}[H]
		\centering
		\begin{tabular}{|l|c|c|c|}
			\hline
			&  \textbf{Global} & \textbf{Smallest local} & \textbf{Biggest local} \\ \hline
			\#nodes   & 5490 &  534 & 4698   \\ \hline
		\end{tabular}
		\caption{Mesh}
		\label{Meshes2}
	\end{table}
	
\begin{table}[H]
	\centering
	\begin{tabular}{|l|C{2.1cm}|C{2.1cm}|} \hline
		Variant& \textbf{Sync.}\newline  Aitken & \textbf{Async.}\newline  $\omega_\mathrm{opt}$ \\\hline
		{Time (s)}   &  881.55   & 79.44  \\ 
		\#iter. glob. &  36   & 506  \\ 
		\#loc. sol. [min, max] &  $\cdot$   &   [348, 6788]  \\ \hline
	\end{tabular}
	\caption{Poor load balancing case: Iterations \& Time (thermal problem)}
	\label{imabalance_linear_thermal_table}
\end{table}

%	\begin{table}[H]
%		\centering
%		\begin{tabular}{|l|c|c|c|c|c|c|}
%			\hline
%			&  \textbf{Aitken} & \textbf{Relaxed asynchronous} \\ \hline
%			Time(s) & 881.55  & \textbf{79.44} \\ \hline
%			\#It. [Loc$_{\min}$,Loc$_{\max}$]  &   36 & 506[348, 6788]  \\ \hline
%		\end{tabular}
%		\caption{Poor load balancing case: Iterations \& Time (thermal problem)}
%		\label{imabalance_linear_thermal_table}
%	\end{table}

\begin{table}[H]
	\centering
	\begin{tabular}{|l|C{2.1cm}|C{2.1cm}|} \hline
		Variant& \textbf{Sync.}\newline  Aitken & \textbf{Async.}\newline  $\omega_\mathrm{opt}$ \\\hline
		{Time (s)}   &  3509.6   & 1904.34  \\ 
		\#iter. glob. &  113   & 2354  \\ 
		\#loc. sol. [min, max] &  $\cdot$   &   [818, 2951]  \\ \hline
	\end{tabular}
\caption{Poor load balancing case: Iterations \& Time (linear elasticity problem)}
\label{imabalance_linear_elasticity_table}
\end{table}

%\begin{table}[H]
%	\centering
%	\begin{tabular}{|l|c|c|c|c|c|c|}
%		\hline
%		&  \textbf{Aitken} & \textbf{Relaxed asynchronous} \\ \hline
%		Time(s) & 3509.6  & \textbf{1904.34} \\ \hline
%		Iterations  &  113  & 2354[818,2951]  \\ \hline
%	\end{tabular}
%	\caption{Poor load balancing case: Iterations \& Time (linear elasticity problem)}
%	\label{imabalance_linear_elasticity_table}
%\end{table}

This case study has been performed using 257 processors, one for the global problem and one processor for each one of the 256 local problems.	
Table \ref{imabalance_linear_thermal_table} and~\ref{imabalance_linear_elasticity_table}  show the computation time and the number of iterations. 

We see that even if the number of iterations can be very large in the asynchronous case, the CPU time is much reduced: 10 times in the thermal case and 2 times for the elasticity case. Again, this highlights the prohibitive cost of synchronization.

%We see, that the computation time is drastically reduced by switching to asynchronous by a coefficient of 10. even if we perform a few iterations in Aitken, highlighting the prohibitive cost of synchronization. These results make us confident that the asynchronous model performs much better than the synchronous one in the case of load imbalance.

%Table \ref{imabalance_linear_elasticity_table} shows the execution time of the asynchronous model and the Aitken one plus the number of iterations performed. The asynchronous model is two times faster than the Aitken, even if the last one performs a few iterations, highlighting the prohibitive cost of synchronization. These results make us confident again that the asynchronous model is more suitable than the synchronous one in the case of load imbalance.